\documentclass[aps,pra, twocolumn,superscriptaddress]{revtex4-2}
\usepackage{amsmath}
\usepackage{bm}
\usepackage{amssymb,amsfonts,latexsym,fancyhdr,graphicx,epstopdf}
\usepackage{graphicx}
\usepackage[english]{babel}
\usepackage{graphicx}
\usepackage {subfigure}
\usepackage[pdftex,colorlinks=true,allcolors=blue]{hyperref}
\usepackage{xcolor}
\usepackage{braket}

\usepackage{feynmp}
\usepackage{amsmath,tkz-euclide}
\def\fisunical{Dip. di Fisica, Universit\`{a} della Calabria, Via P. Bucci, Cubo 30C, I-87036 Rende (CS), Italy}
\def\infnlnfcs{INFN, sezione LNF, Gruppo collegato di Cosenza, Via P. Bucci, Cubo 31C, I-87036 Rende (CS), Italy}
\def\fisYork{School of Physics, Engineering and Technology, The University of York, YO10 5DD, York, United Kingdom}
\DeclareMathOperator{\Tr}{Tr}
\newcommand{\msc}[1]{\normalfont\textsc{#1}}
\begin{document}

\title{Thermal density functional theory approach to quantum thermodynamics}

\author{\firstname{Antonio}~\surname{Palamara}}
\affiliation{\fisunical}
\affiliation{\infnlnfcs}

\author{\firstname{Francesco}~\surname{Plastina}}
\affiliation{\fisunical}
\affiliation{\infnlnfcs}

\author{\firstname{Antonello}~\surname{Sindona}}
\affiliation{\fisunical}
\affiliation{\infnlnfcs}

\author{\firstname{Irene}~\surname{D'Amico}}
\email{corresponding author: irene.damico@york.ac.uk}
\affiliation{\fisYork}

\begin{abstract}
Understanding the thermodynamic properties of many-body quantum systems and their emergence from microscopic laws is a topic of great significance due to its profound fundamental implications and extensive practical applications. 
Recent advances in experimental techniques for controlling and preparing these systems have increased interest in this area, as they have the potential to drive the development of quantum technologies. 
In this study, we present a density-functional theory approach to extract detailed information about the statistics of work and the irreversible entropy associated with quantum quenches at finite temperature. 
Specifically, we demonstrate that these quantities can be expressed as functionals of thermal and out-of-equilibrium densities, which may serve as fundamental variables for understanding finite-temperature many-body processes. We, then, apply our method to the case of the inhomogeneous Hubbard model, showing that our density functional theory based approach can be usefully employed to unveil the distinctive roles of interaction and external potential on the thermodynamic properties of such a system.
\end{abstract}

\maketitle
\renewcommand\thesubsection{\thesection.\arabic{subsection}}
\renewcommand\thesubsubsection{\thesubsection.\arabic{subsubsection}}
\makeatletter
\renewcommand{\p@subsection}{}
\renewcommand{\p@subsubsection}{}
\makeatother

\section{Introduction} 
Density-functional theory~(DFT)~\cite{hohenberg1964density,kohn1965self} and its time-dependent~(TD) extension~\cite{runge1984density,ullrich:2012} are powerful and well-established methods for studying the electronic properties of interacting many-body systems at zero temperature, with DFT providing comprehensive access to ground state properties and TDDFT extending this capability to include the prediction of excited states.
Thermal density-functional theory~(ThDFT), introduced by Mermin~\cite{mermin1965thermal}, extends the Hohenberg-Kohn~(HK) framework of DFT~\cite{hohenberg1964density} to address the electronic properties of many-body systems under conditions where accounting for finite temperature effects is indispensable~\cite{BSGP16,PGB16,SPB16,YTPB14,PPFS11,PhysRevB.105.235114,PhysRevLett.125.055002,PhysRevB.94.241103,Karasiev:2016,Trickey:2014}.

Renewed attention in ThDFT has been driven by the study of thermal properties of out-of-equilibrium interacting quantum systems, which represents a major focus in quantum thermodynamics~(QT).
This field is rapidly evolving thanks to advances in preparing and coherently controlling quantum systems at the microscopic scale, enabling experimental verification of fundamental properties such as fluctuation theorems~\cite{PhysRevLett.78.2690,RevModPhys.81.1665,PhysRevLett.113.140601,PhysRevLett.115.190601}. 
These developments also hold potential for new quantum technologies based on complex quantum systems (see e.g. \cite{campaioli2024colloquium}).
A major goal in QT is to understand the role of purely quantum features, such as coherence and correlations, in thermodynamic processes.
This notably includes work processes, where work is extracted from or performed on a quantum system, and the generation of irreversible entropy~\cite{campisi2011quantum,goold2016role,klatzow2019experimental,perarnau2015extractable,korzekwa2016extraction, shi2022entanglement,gour2022role,rodrigues2024nonequilibrium,onishchenko2024probing,PhysRevE.99.042105,PhysRevE.105.014101,PhysRevResearch.2.033279,varizi2021contributions,PhysRevResearch.5.043104,zawadzki2024work,francica2020quantum,francica2017daemonic,PRXQuantum.4.020353}.

In this realm, the emphasis is rapidly shifting towards coupled many-body systems; see, e.g, ~\cite{geier2022non,hahn2023quantum}. Indeed, recent studies have demonstrated that particle interactions can enhance the efficiency of quantum heat engines~\cite{zhang2024energy,jussiau2023many,PhysRevResearch.5.043104,jaramillo2016quantum,PhysRevResearch.2.032062}.
On the theoretical side, addressing finite-temperature quantum many-body systems poses significant challenges, often requiring approximations to manage their complexity.

In this context, and drawing inspiration from previous works~\cite{zawadzki2022approximating,skelt2019many,herrera2018melting,herrera2017dft}, our objective is to establish a robust theoretical framework for applying DFT to the study of non-equilibrium thermodynamics in quenched interacting many-body systems. Our formalism focuses on the canonical ensemble, of particular relevance for QT thermal machines.
Specifically, we demonstrate that the thermal and out-of-equilibrium densities form the basis of an ab initio framework for deriving thermodynamic properties of quantum systems that experience a sudden quench.
A key advantage is that the out-of-equilibrium thermodynamics of interacting many-body systems can be effectively investigated using the Kohn-Sham~(KS) approach to DFT~\cite{kohn1965self}. 
The strength of this approach lies in its ability to evaluate, in principle exactly, the thermal densities by mapping the original interacting many-body system onto a fictitious non-interacting one.
We thus establish a general framework for the DFT approach to QT, specifically for the canonical ensemble, and validate it through the analysis of the quenched inhomogeneous Hubbard model. 
Accordingly, we first compare results from exact diagonalization for small systems with those obtained using our finite temperature KS  mapping, and then extend our analysis to larger systems. 
This extension allows us to clarify how interactions influence thermal densities and, consequently, the work performed.

The reminder of the paper is organized as follows. 
In Sec.~\ref{sec2}, we recall the Mermin-Hohenberg-Kohn~(MHK) theorem~\cite{hohenberg1964density,mermin1965thermal}, adapting it to the context of lattice Hamiltonians in closed quantum systems. 
In Sec.~\ref{sec3}, we review the fundamental concepts of out-of-equilibrium QT, showing that for work protocols of infinitesimal duration, the probability distributions of work and irreversible entropy production are completely determined by the finite-temperature equilibrium densities of the pre-quench and post-quench Hamiltonians. 
In Sec.~\ref{sec4}, we recall the Mermin-Kohn-Sham~(MKS) equations~\cite{kohn1965self,mermin1965thermal}, employing the finite-temperature KS mapping to calculate the thermodynamic quantities of interest. 
In Sec.~\ref{sec5}, we present a method for calculating the thermal densities of a system of indistinguishable particles within the canonical ensemble using the KS mapping.
In Sec.~\ref{sec6}, we apply our theoretical framework to short Hubbard chains, solving the problem both exactly and via the finite-temperature KS mapping to validate the robustness and accuracy of our DFT-based approach. 
We then extend our DFT-based analysis to the Hubbard model with a larger number of sites. Finally, we draw our conclusions in Sec. \ref{sec7}.

\section{Thermal \textbf{\{A\}}-functional theory\label{sec2}}
In its original formulation, DFT is founded on the HK theorems~\cite{hohenberg1964density}, which were later generalized to finite temperatures by Mermin~\cite{mermin1965thermal}.
These theorems establish a one-to-one correspondence between the external potential $v(\mathbf{r})$~(or $v(\mathbf{r}) - \mu$, where $\mu$ is the chemical potential), the quantum (equilibrium) state of the system, and the ground state~(or thermal) electron density $n(\mathbf{r})$~(or $n^{\beta}(\mathbf{r})$).
More recently, it has been demonstrated that the HK theorems at zero temperature are still valid for quantum systems whose Hamiltonian is defined on a lattice~\cite{PhysRevA.74.052335,PhysRevB.52.2504,coe2015uniqueness}, with some limitations~\cite{penz2021density,xu2022extensibility,francca2018testing}.
Similarly, it can be shown that this extension also holds true for finite-temperature closed quantum systems. 

To demonstrate this, let us consider a closed quantum system governed by a Hamiltonian $\hat{\mathcal{H}}$ of the following form:
\begin{equation}
\hat{\mathcal{H}}[\{\lambda_i\}]= \hat{\mathcal{H}}_{0}+ \hat{\mathcal{V}}_{\msc{ext}}[\{\lambda_i\}]
=\hat{\mathcal{H}}_{0} + \sum^{\mathcal{L}}_{i=1}\lambda_i \hat{A}_i,
\label{Hl}
\end{equation}
where $\hat{\mathcal{H}}$ is quite general and can describe spin chains, fermionic, or bosonic systems. 
$\hat{\mathcal{H}}_{0}$ is the `\textit{universal}' HK Hamiltonian, and $\hat{\mathcal{V}}_{\msc{ext}}$ is the external potential, controlled by a set of $\mathcal{L}$ parameters $\{\lambda_i\}$ that multiply the local operators $\{\hat{A}_i\}$.

Let us denote the thermal expectation value of $\hat{A}_i$ as 
\begin{equation} 
a^\beta_i := \Tr\{\hat{\rho}^{\msc{th}}_\beta[\{\lambda_i\}]\hat{A}_i\},
\label{tha}
\end{equation}
where
$\hat{\rho}^{\msc{th}}_\beta[\{\lambda_i\}]= 
{e^{-\beta \hat{\mathcal{H}}[\{\lambda_i\}]}}/{\mathcal{Z}[\{\lambda_i\}]}
$
represents the Gibbs state and $\mathcal{Z}[\{\lambda_i\}] = \Tr\{e^{-\beta \hat{\mathcal{H}}[\{\lambda_i\}]}\}$ expresses the canonical partition function.
By definition, each $a^\beta_i$ is a function of the parameters $\{\lambda_i\}$.
It can be shown that exactly one parameter set $\{\lambda_i\}$ corresponds to a given mean value set $\{a^\beta_i\}$~(see appendix \ref{app1}).

Since $\{a^\beta_i\}$ uniquely determines $\{\lambda_i\}$, which in turn determines $\hat{\rho}^{\msc{th}}_\beta$, the thermal state is also a unique functional of $\{a^\beta_i\}$.
This relationship highlights the direct connection between observable mean values and the underlying thermal state, emphasizing the role of $\{a^\beta_i\}$ as the fundamental descriptors of the system:
\begin{equation}
 \{\lambda_i\} \iff \{a^\beta_i\} \iff \hat{\rho}^{\msc{th}}_\beta \equiv \hat{\rho}^{\msc{th}}_\beta[\{a^\beta_i\}].
\label{mainn}
\end{equation}

By the one-to-one relations established in Eq.~(\ref{mainn}), the thermal HK theorem, as demonstrated by Mermin~\cite{mermin1965thermal}, remains valid for closed quantum systems defined by the  Hamiltonian~(\ref{Hl}). 
Consequently, the free energy 
\begin{equation}
\label{F(rho)}
 \mathcal{F}[\hat{\rho}]:=\Tr\left\{\hat{\rho}\left(\hat{\mathcal{H}}+\frac{\ln{\hat{\rho}}}{\beta}\right)\right\},
 \end{equation}
minimized by the equilibrium Gibbs state, can be expressed as a unique function of $\{\lambda_i\}$ or, equivalently, as a unique functional of $\{a^\beta_i\}$:
\begin{equation}
\mathcal{F}[\{a^\beta_i\}] = \Omega[\{a^\beta_i\}]+ \sum^{\mathcal{L}}_{i=1} \lambda_i a^\beta_i. 
 \label{freeen}
\end{equation}
Here, $\Omega[\{a^\beta_i\}]$ represents the generally unknown `\textit{universal}'  functional~\cite{dreizler2012density}, associated with the $\{\lambda_i\}$-independent Hamiltonian $\hat{\mathcal{H}}_0$.
Similarly to the zero-temperature case~\cite{PhysRevA.74.052335,PhysRevB.52.2504}, Eq.~(\ref{freeen}) allows us to interpret the HK theorem as an expression of duality in the sense of a Legendre transform. 
This duality relates the thermal expectation values $\{a^\beta_i\}$, which play a role analogous to the thermal electron densities $n^{\beta}(\mathbf{r})$, to the set of work parameters $\{\lambda_i\}$, which fulfill a role analogous to the external potentials $v(\mathbf{r})$.

A special case of the Hamiltonian~(\ref{Hl}) corresponds to the following single-parameter scenario:
\begin{equation}
\hat{\mathcal{H}}[\lambda]= \hat{\mathcal{H}}_{0}+ \lambda\sum^{\mathcal{L}}_{i=1} \hat{A}_i,
\label{Hll}
\end{equation}
where the HK theorem remains valid. 
However, to ensure a unique mapping between $\lambda$ and 
\begin{equation}
\braket{\hat{A}} \equiv \Tr\bigg\{\hat{\rho}^{\msc{th}}_\beta[\{\lambda_i\}] \sum_{i=1}^{\mathcal{L}} \hat{A}_i\bigg\},
\end{equation}
it is necessary to require that $[\hat{\mathcal{H}}_{0}, \sum_{i=1}^{\mathcal{L}} \hat{A}_i] \neq 0$~\cite{alcaraz2008finite}.

Another important tool is offered by the Hellmann-Feynman~(HF) theorem~\cite{PhysRevA.74.052335,pons2020hellmann,landi2021irreversible},  
which establishes a relationship between the first derivative of the free energy with respect to the $i$-th external parameter and the $i$-th thermal density:
\begin{equation}
\frac{\partial \mathcal{F}}{\partial \lambda_i} = \Tr\bigg\{{ \hat{\rho}^{\msc{th}}_\beta[\{\lambda_i\}] \frac{\partial \hat{\mathcal{H}}}{\partial \lambda_i}}\bigg\} 
=\Tr\{ \hat{\rho}^{\msc{th}}_\beta[\{\lambda_i\}]\hat{A}_i\}=a^\beta_i.
\label{af}
\end{equation}
This equation will be especially useful in the subsequent sections, particularly in the context of thermal sudden quenches, where thermal expectation values serve as fundamental variables for deriving the thermodynamic quantities of interest.

\section{Thermal \textbf{\{A\}}-functional theory approach to quantum thermodynamics\label{sec3}} 
ThDFT can be effectively employed to extract information about the out-of-equilibrium thermodynamics of a closed quantum system by leveraging its ability to handle thermal and quantum fluctuations.
To establish the formalism, we first review some key concepts in QT.
Our focus is on a closed quantum system that has been driven out of equilibrium by a unitary quantum process.  
Specifically, we consider a generic closed quantum system characterized by the Hamiltonian~(\ref{Hl}), which depends on a set of time-dependent work parameters $\{\lambda^t_i\}$. 
The system is initially in equilibrium with a bath at inverse temperature $\beta$. 
In this configuration, at $t=0$, the set of work parameters $\{\lambda^0_i\}$ defines the thermal state, 
represented by the Gibbs density operator $\hat{\rho}^{\msc{th}}_\beta[\{\lambda^0_i\}]$. 
After the system is decoupled from the bath, it undergoes unitary dynamics, from $t=0$ to $t=\tau$,  governed by the time evolution operator $\hat{\mathcal{U}}(\tau,0)$. 
This evolution is driven by a protocol that changes the set of work parameters from $\{\lambda^0_i\}$, with corresponding mean values $\{a^{\beta{\,}0}_i\}$,  to $\{\lambda^f_i\}$, with corresponding mean values $\{a^{\beta{\,}f}_i\}$, over the finite time interval $\tau$.

\subsection{Probability distributions of work and irreversible entropy production}
The work performed or extracted during the protocol is not an observable and cannot be represented by a Hermitian operator. Rather, work is a stochastic variable characterized by a probability distribution, which is determined by performing two projective measurements at the initial and final times, respectively~\cite{talkner2007fluctuation}. 
These measurements involve the instantaneous eigenbasis of the system Hamiltonian
denoted $\{\epsilon_n(t), \ket{n(t)}\}$.
The probability distribution of work~(PDW) is defined as follows:
\begin{equation}
P_\tau(w)=\sum_{nm}p_{n}(0)p_{n \rightarrow m}(\tau)\delta(w-w_{mn}),
\label{pw}
\end{equation}
where $w_{mn} = \epsilon_m(\tau) - \epsilon_n(0)$ is the work performed in a single realization, $p_n(0) \equiv \bra{n(0)} \hat{\rho}^{\msc{th}}_\beta \ket{n(0)}$ is the probability of finding the system in the $n$-th eigenstate at time $t=0$, and $p_{m \rightarrow n}(\tau)$ is the transition probability, between the $n$-th and $m$-th eigenstates, due to the protocol. 
Fluctuations arising from the protocol and the measurements are encoded by $p_n(0)p_{m \rightarrow n}(\tau)$ and constrained by the Jarzynski's equality~\cite{PhysRevLett.78.2690} 
\begin{equation}
\braket{e^{-\beta w}} = e^{-\beta\Delta\mathcal{F}},
\label{J}
\end{equation}
where  $\Delta\mathcal{F}= \mathcal{F}[\{\lambda^f_i\}]-\mathcal{F}[\{\lambda^0_i\}]$ is the free energy difference between the two equilibrium configurations, corresponding to the initial and final Hamiltonian.

By the Jensen's inequality, Eq.~(\ref{J}) implies that $\braket{w} \geq \Delta \mathcal{F}$, which reflects the second law of thermodynamics. 
This leads to the definition of the average irreversible work~\cite{crooks1999entropy}
\begin{equation}
\braket{w_{\msc{irr}}} := \braket{w} - \Delta \mathcal{F},
\end{equation}
which is directly related to the average  irreversible entropy production~\cite{deffner2010generalized, kawai2007dissipation, 
talkner2007fluctuation}
\begin{equation}
\braket{\mathcal{S}_{\msc{irr}}}:=\beta\braket{w_{\msc{irr}}}.
\label{sirr}
\end{equation}
Both $\braket{w_{\msc{irr}}}$ and $\braket{\mathcal{S}_{\msc{irr}}}$ give a measure of the irreversibility introduced by performing the unitary transformation  $\hat{\rho}(\tau)=\hat{\mathcal{U}}(\tau,0)\hat{\rho}^{\msc{th}}_\beta[\{\lambda^0_i\}]\hat{\mathcal{U}}^\dagger(\tau,0)$.
Strictly speaking, due to the unitary nature of the time evolution, no von Neumann entropy is generated during this process, with the entropy of the system remaining constant:
\begin{equation}
\mathcal{S}(\hat{\rho}(\tau))=-\Tr\{\hat{\rho}(\tau)\log{\hat{\rho}(\tau)}\}\equiv \mathcal{S}(\hat{\rho}^{\msc{th}}_\beta[\{\lambda^0_i\}]).
\end{equation}
Equation~(\ref{sirr}) is referred to as a measure of irreversibility because, when the system is returned to the bath after the protocol, it relaxes from the out-of-equilibrium state to the thermal state $\hat{\rho}^{\msc{th}}_\beta[\{\lambda^f_i\}]$. 
This relaxation is a non-unitary process, and the entropy produced during this process is precisely $\braket{\mathcal{S}_{\msc{irr}}}$.

We emphasize that, like the average work, $\braket{\mathcal{S}_{\msc{irr}}}$ is also the first moment of a probability distribution obtained within the two-point measurement framework.
Indeed, it is possible to define a stochastic variable,  associated with the production of irreversible entropy, as follows:  
\begin{equation}
s_{mn}:=\beta(\epsilon_m(\tau)-\epsilon_n(0))-\beta\Delta\mathcal{F}.
\end{equation}
Then, the probability distribution for entropy production~(PDE), analogous to the PDW, takes the form
\begin{equation}
P_\tau(s)=\sum_{nm}p_n(0)p_{n \rightarrow m}(\tau)\delta(s-s_{mn}).
\label{ps}
\end{equation}

A complementary approach to investigate the statistical properties of work processes and irreversible entropy production is based on the Fourier transforms of the corresponding probability distributions, namely, $P_\tau(w)$ from Eq.~(\ref{pw}) and $P_\tau(s)$ from Eq.~(\ref{ps}). 
It is therefore convenient to rely on the characteristic function of work~\cite{talkner2007fluctuation}
\begin{align}
\label{chiw}
\chi_\nu(w,\tau) :=& \int dw e^{i\nu w}P_\tau(w)   \\
 =&\Tr\{
e^{i\nu \hat{\mathcal{H}}[\{\lambda^f_i\}]}
\hat{\mathcal{U}}(\tau,0)
e^{-i\nu \hat{\mathcal{H}}[\{\lambda^0_i\}]}\notag \\
&\qquad\qquad\quad\times\hat{\rho}^{\msc{th}}_\beta([\{\lambda^0_i\}])
\hat{\mathcal{U}}^\dagger(\tau,0)\},\notag 
\end{align}
with associated moments 
\begin{equation}
\braket{w^n(\tau)} = (-i)^n\partial^n_\nu \chi_\nu(w,\tau)|_{\nu=0}.
\label{dchi}
\end{equation} 
It is further instructive to introduce the characteristic function for irreversible entropy production:
\begin{align}
\label{chis}
\chi_\mu(s,\tau) :=& \int ds e^{i\mu s} P_\tau(s) \\
=& e^{-i \beta \mu \Delta\mathcal{F}} 
\Tr\{
e^{i\beta \mu \hat{\mathcal{H}}[\{\lambda^f_i\}]} \hat{\mathcal{U}}(\tau,0) \notag
\\
&\qquad\times
e^{-i\beta \mu \hat{\mathcal{H}}[\{\lambda^0_i\}]}
\hat{\rho}^{\msc{th}}_\beta[\{\lambda^0_i\}]
\hat{\mathcal{U}}^\dagger(\tau,0)
\},\notag
\end{align}
with associated moments
\begin{equation}
\braket{s^n(\tau)} = (-i)^n\partial^n_\mu \chi_s(w,\tau)|_{\mu=0}.
\label{dchis}
\end{equation} 
The two quantities expressed in Eqs.~(\ref{chiw}) and~(\ref{chis}) play a crucial role in the development of the thermal density functional framework for specific protocols, as detailed in Sec.~\ref{sub:sqp} (sudden quench) and appendix \ref{app2} (finite-time protocols).

\subsection{The sudden quench protocol\label{sub:sqp}}
A sudden quench involves an instantaneous shift of the work parameters, from $\{\lambda^0_i\}$ to $\{\lambda^f_i\}$.
This variation occurs in an infinitesimally short timeframe, unlike the finite-time protocols covered in appendix \ref{app2}. 
It can be demonstrated that all of the moments of the PDW in this scenario are functionals of the initial thermal densities.
This outcome is due to the fact that, in a sudden quench, the time  evolution operator approaches the identity operator, as the quench duration becomes infinitesimally small~\cite{messiah2014quantum}:  $\lim\limits_{\tau\rightarrow 0^{+}}\hat{\mathcal{U}}(\tau,0)=\hat{1}$. 
Therefore, the characteristic function of work~(\ref{chiw}) simplifies to an ensemble average over the initial Gibbs state:
\begin{equation}
\chi_\nu(w,0^+) = \Tr\{
e^{i\nu \hat{\mathcal{H}}[\{\lambda^f_i\}]}
e^{-i\nu \hat{\mathcal{H}}[\{\lambda^0_i\}]}
\hat{\rho}^{\msc{th}}_\beta[\{\lambda^0_i\}]\}.
\label{chiws}
\end{equation}
For a Hamiltonian of the form~(\ref{Hl}), any thermal average over the initial state can be expressed as a functional of the initial mean values $\{a^{\beta{\,}0}_i\}$, as dictated by the generalized HK theorem outlined in Sec.~\ref{sec2} and appendix \ref{app1}.
Consequently, we have: $\chi_\nu(w,0^+)=\chi_\nu(w,0^+)[\{a^{\beta{\,}0}_i\}]$, which implies that all the moments of $P_{0^+}(w)$ 
are functionals of $\{a^{\beta{\,}0}_i\}$, with parametric dependence on both $\{\lambda^0_i\}$ and $\{\lambda^f_i\}$.
More explicitly, by using Eq.~(\ref{dchi}), these moments can be expressed as the following thermal averages:
\begin{equation}
\braket{w^n}= \Tr\{(\hat{\mathcal{H}}[\{\lambda^f_i\}]-\hat{\mathcal{H}}[\{\lambda^0_i\}])^n\hat{\rho}^{\msc{th}}_\beta[\{\lambda^0_i\}]\}, \label{w^n}
\end{equation}
which are functional of the initial thermal densities: $\braket{w^n}=\braket{w^n}[\{a^{\beta{\,}0}_i\}]$.
Now, the linearity of the Hamiltonian $\mathcal{H}[\{\lambda_i\}]$ in the work parameters $\{\lambda_i\}$ leads to 
\begin{equation}
\hat{\mathcal{H}}[\{\lambda^f_i\}]-\hat{\mathcal{H}}[\{\lambda^0_i\}]=\sum_i(\lambda^f_i-\lambda^0_i)\frac{\partial \hat{\mathcal{H}}}{\partial \lambda^0_i}.  
\label{line}
\end{equation}
Then, using Eqs.~(\ref{w^n}) and~(\ref{line}), 
the average work becomes:
\begin{equation}
\braket{w}=\sum_i(\lambda^f_i-\lambda^0_i)a^{\beta{\,}0}_i
\label{ws}
\end{equation}

Similarly, the characteristic function of the irreversible entropy production~(\ref{ps}),  for a sudden quench protocol, takes the simplified expression
\begin{align}
\chi_\mu(s,0^+)=&e^{-i \beta \mu \Delta\mathcal{F}}\Tr\{
e^{i\beta \mu \hat{\mathcal{H}}[\{\lambda^f_i\}]} \notag \\
&\qquad\qquad\times e^{-i\beta \mu \hat{\mathcal{H}}[\{\lambda^0_i\}]}
\hat{\rho}^{\msc{th}}_\beta[\{\lambda^0_i\}]
\}.
\label{chiss}
\end{align}
Again, by virtue of the thermal HK theorem, the final and initial free energies in $\Delta\mathcal{F}$ are functionals of $\{a^{\beta{\,}f}_i\}$ and $\{a^{\beta{\,}0}_i\}$, respectively.
On the other hand, as seen in Eq.~(\ref{chiws}), the trace in Eq.~(\ref{chiss}) is a functional of $\{a^{\beta{\,}0}_i\}$ only. 
Consequently, we can assert that $P_{0^+}(s)$, or $\chi_\mu(s,0^+)$, and the associated moments $\braket{s^n}$ are functionals of both $\{a^{\beta{\,}f}_i\}$ and $\{a^{\beta{\,}0}_i\}$.
In particular, the average irreversible entropy production takes the form:
\begin{align}
\label{Ss}
\braket{\mathcal{S}_{\msc{irr}}}=&\beta\sum_i(\lambda^{f}_{i}-\lambda^{0}_{i})a^{\beta{\,}0}_i\\
&\qquad-\beta\{\mathcal{F}[\{a^{\beta{\,}f}_i\}]-\mathcal{F}[\{a^{\beta{\,}0}_i\}]\} \notag
\end{align}

We focus on sudden quenches of infinitesimal variation, where the work parameters $\{\lambda^0_i\}$ change by an elementary amount to $\{\lambda^0_i + \delta \lambda_i\}$. 
In this context, we can derive an explicit functional form for the average irreversible entropy production.
In particular, we can expand Eq.~(\ref{Ss}) in a Taylor series and apply the HF theorem, as expressed by Eq.~(\ref{af}). 
This yields:
\begin{equation}
\braket{\mathcal{S}_{\msc{irr}}}  = -\frac{\beta}{2}\sum_{i,j}\delta\lambda_i\delta\lambda_j\frac{\partial a^{\beta{\,}0}_i}{\partial \lambda^0_j}.
\label{Sis}
\end{equation}
We emphasize that Eqs.~(\ref{ws}) and~(\ref{Sis}) demonstrate that the mean values of work and irreversible entropy production are explicit functionals of the initial thermal densities.  
As discussed in the following sections, this is particularly important for extracting information about many-body systems using the KS mapping, which enables the computation of thermal electron densities within a formally non-interacting framework.

\subsection{Fluctuation-dissipation relations in the sudden quench limit\label{sub:fdr}}
We now recall that classical quasi-adiabatic processes follow the fluctuation-dissipation relation~(FDR) 
\begin{equation}
\braket{\mathcal{S}_{\msc{irr}}}=\frac{\beta^2}{2}\sigma^2_w,
\label{FDRcl}
\end{equation}
where $\sigma^2_w=\braket{w^2}-\braket{w}^2$ represents the variance in the PDW~\cite{wood1991systematic,hendrix2001fast,PhysRevLett.78.2690}.
Recently, it has been demonstrated that for slow processes in open quantum systems, close to equilibrium, the FDR is given by Eq.~(\ref{FDRcl}) minus a positive, purely quantum  term, which arises from the non-commutativity of the thermodynamic protocol~\cite{miller2019work,scandi2020quantum}. 
Here, with the aim of obtaining an explicit functional form in terms of initial thermal densities for the second moment of the work probability distribution, we reobtain a similar generalized FDR  that holds in the infinitesimal sudden quench regime.
This should not be surprising, as an adiabatic process, i.e., one that is close to equilibrium throughout, can be considered as a sequence of a large number of sudden quenches, each followed by thermalization towards the equilibrium state \cite{scandi2020quantum}.\\

To this end, we focus on the second moment of the PDW. 
Then, we distinguish the case where the final and initial Hamiltonians share a common eigenbasis, and the case where they do not.
Additional details on the following derivations are provided in appendix \ref{app3}.
In the specific scenario where $[\hat{\mathcal{H}}[\{\lambda^f_i\}],\hat{\mathcal{H}}[\{\lambda^0_i\}]]=0$, the second moment of the PDW is given by:
\begin{equation}
\braket{w^2}_c=\sum_{i,j}\delta\lambda_i\delta\lambda_j a^{\beta{\,}0}_ia^{\beta{\,}0}_j-\frac{1}{\beta}\sum_{i,j}\delta\lambda_i\delta\lambda_j\frac{\partial a^{\beta{\,}0}_i }{\partial \lambda^0_j}.
\label{w^2}
\end{equation}
Notably, Eq.~(\ref{w^2}) expresses an explicit functional of the initial mean values, independently of the amplitude of the sudden  quench.
Nonetheless, with an infinitesimal sudden quench, we can utilize the expressions for the average work, Eq.~(\ref{ws}), and the average irreversible entropy production, Eq.~(\ref{Sis}), to rewrite Eq.~(\ref{w^2}) as: 
\begin{equation}
\braket{w^2}_c=\braket{w}^2+\frac{2}{\beta^2}\braket{\mathcal{S}_{\msc{irr}}},
\end{equation}
This relation validates the FDR, in its classical form, as given by Eq.~(\ref{FDRcl}), 
to the leading order in $\{\delta\lambda_i\}$.

Turning to the instance where the initial and final Hamiltonians do not commute, and using Eq.~(\ref{w^n}), the second moment of the PDW is still a functional of the initial equilibrium thermal densities.
Specifically, the latter can be split into the following two parts: 
\begin{equation}
\braket{w^2}=\braket{w^2}_c+ \Theta_2[\{a^{\beta{\,}0}_i\}], 
\label{w^22}
\end{equation}
where $\Theta_2[\{a^{\beta{\,}0}_i\}]$ arises directly from the incompatibility of the two Hamiltonians. 
A possible approximation method for this functional is provided in sec \ref{sec4}.
Operating again in the infinitesimal sudden quench limit, we can plug Eqs.~(\ref{ws}) and~(\ref{Sis}) into Eq.~(\ref{w^22}).
By doing so, we recover the generalized FDR,
\begin{equation}
\braket{\mathcal{S}_{\msc{irr}}}=\frac{\beta^2}{2}\sigma^2_w -\frac{\beta^2}{2}\Theta_2[\{a^{\beta{\,}0}_i\}],
\label{FDR2}
\end{equation}
which takes into account both thermal fluctuations and quantum fluctuations due to $[\hat{\mathcal{H}}[\{\lambda^f_i\}],\hat{\mathcal{H}}[\{\lambda^0_i\}]]\neq0$.

\section{Thermal Kohn-Sham mapping for quantum thermodynamics\label{sec4}}  
The results discussed so far in previous sections are formally exact, at least in the limiting conditions of the protocols investigated for the evolution of the coupling parameters.
However, as a many-body system grows in complexity, the number of interactions and possible configurations needed to determine the exact thermal density becomes computationally infeasible.  
To address this challenge, the KS scheme~\cite{kohn1965self} provides a powerful approach within the framework of DFT for developing efficient approximations. 
This method relies on defining a formally non-interacting many-body system, the KS system, which is designed to replicate the same particle density as the original interacting physical system.

For systems governed by the Hamiltonian~(\ref{Hl}), and building on methods developed in earlier studies~\cite{PhysRevB.52.2504} at zero temperature, the KS approach can be applied as follows.
We assume the existence of a set of auxiliary systems, each described by the Hamiltonian 
\begin{equation}
\hat{{\mathcal{H}}}^{\msc{ks}}=\hat{{\mathcal{H}}}^{\msc{ks}}_{0}+ \sum^{\mathcal{L}}_{i=1}{\lambda}^{\msc{ks}}_i \hat{A}_i,
\end{equation}
where the one-body operator $\hat{\mathcal{H}}^{\msc{ks}}_{0}$ replaces the complex many-body term $\hat{\mathcal{H}}_{0}$ in Eq.~(\ref{Hl}).
The KS Hamiltonian simplifies the problem by focusing on non-interacting particles in an effective potential associated to specific coupling parameters.
We further assume that $\hat{\mathcal{H}}^{\msc{ks}}$ yields the same set of thermal densities as the original Hamiltonian:
\begin{equation}
\Tr\{\hat{\rho}^{\msc{th}}_\beta[\{\lambda_i\}]\hat{A}_i\}=\Tr\{\hat{{\rho}}^{\msc{th}}_{\beta{\msc{ks}}}[\{{\lambda}^{\msc{ks}}_i\}]\hat{A}_i\}.
\end{equation}
In this setting, $\hat{\rho}^{\msc{th}}_\beta$ and $\hat{\rho}^{\msc{th}}_{\beta, \msc{ks}}$ denote the thermal density matrices of the original and KS systems, respectively, both parameterized by inverse temperature $\beta$ and coupling parameters $\{\lambda_i\}$ and $\{\lambda^{\msc{ks}}_i\}$.

The MHK theorem clearly holds for the KS Hamiltonian, though with some restrictions at absolute zero temperature~\cite{PhysRevA.74.052335,PhysRevB.52.2504,coe2015uniqueness}. 
Consequently, the coupling parameters $\lambda^{\msc{ks}}_i$ are functionals of the thermal averages $\{a^\beta_i\}$, i.e., $\lambda^{\msc{ks}}_i = \lambda^{\msc{ks}}_i[\{a^\beta_i\}]$. 
At this point, the following MKS equations can be solved self-consistently for $\{a^\beta_i\}$:
\begin{align}
&\hat{{\mathcal{H}}}^{\msc{ks}}_{0}+\sum^{\mathcal{L}}_{i=1}({\lambda}^{\msc{h-xc}}_i[\{a^\beta_i\}]+\lambda_i)\hat{A}_i\ket{\phi^{i}_{\beta}}=\epsilon^{i}_{\beta}\ket{\phi^{i}_{\beta}} \label{KS1}\\
&a^\beta_i=\Tr\{\hat{{\rho}}^{\msc{th}}_{\beta{\msc{ks}}}[\{\lambda^{\msc{h-xc}}_i[\{a^\beta_i\}]+\lambda_i\}]\hat{A}_i\}.
\label{KS}
\end{align}
Here, the effective parameters $\lambda^{\msc{h-xc}}_i[\{a^\beta_i\}]=\lambda^{\msc{ks}}_i-\lambda_i$ play the role of the Hartree~(H) and exchange-correlation~(XC) potentials in the usual KS mapping,  which account for the effect of the many-body interaction term in $\hat{\mathcal{H}}_0$. 
It is worth recalling that while the eigensystem of the non-interacting KS Hamiltonian exactly reproduces the thermal density, it generally do not correspond to the eigensystem of the interacting Hamiltonian~\cite{capelle2013density}.

The approach outlined here is particularly useful when the one-body Hamiltonian $\hat{\mathcal{H}}^{\msc{ks}}_{0}$ has a simple form, such as in the case of a chain of interacting fermions, where $\hat{\mathcal{H}}^{\msc{ks}}_{0}$ reduces to a kinetic energy operator.
In these scenarios, as is typically done within the KS framework, suitable approximations can be employed for the functionals $\lambda^{\msc{h-xc}}_i[\{a^\beta_i\}]$.
In other terms, any thermodynamic quantity expressed as an explicit functional of the thermal densities can be evaluated through a KS mapping, with an accuracy dictated by the approximations made for the functionals $\lambda^{\msc{h-xc}}_i[\{a^\beta_i\}]$.
Nonetheless, not all quantities in the MKS equations can be directly expressed as functionals of the densities. 
For example, the functional form of $\Theta_2[\{a^{\beta\,0}_i\}]$ in Eq.~(\ref{w^22}) requires reasonable approximations to be determined.

\subsection{Local density approximation for $\Theta_2[\{a^{\beta{\,}0}_i\}]$}
The local density approximation~(LDA) is the simplest and most widely used approach for modeling XC effects in DFT. 
For instance, the LDA has been effectively employed  to develop functionals for calculating the entanglement in spatially inhomogeneous many-fermion systems~\cite{franca2008entanglement}.
To construct an LDA scheme for an inhomogeneous system, it is necessary to have an analytical solution for the corresponding homogeneous problem, where all coupling parameters are equal, i.e., $\lambda_i = \lambda$.  
In the homogeneous case, the functional $\Theta_2[\{a^{\beta{\,}0}_i\}]$ reduces to $\Theta^{\msc{hom}}_2[a^{\beta{\,}0}]$, where:
\begin{equation}
a^{\beta{\,}0}=\frac{1}{{\mathcal{L}}}\Tr\{\hat{\rho}^{\msc{th}}_\beta[\{\lambda_i\}]\sum^{\mathcal{L}}_{i=1}\hat{A}_i\}.
\end{equation}
Based on this, the following LDA scheme can be put forward:
\begin{equation}
\label{theta2lda}
\Theta^{\msc{lda}}_2[\{a^{\beta{\,}0}_i\}]=\sum_i\Theta^{\msc{hom}}_2[a^{\beta{\,}0}]|_{a^{\beta{\,}0}\rightarrow a^{\beta{\,}0}_i}.
\end{equation}  
A crucial aspect of this implementation is that Eq.~(\ref{theta2lda}) approximates the fluctuations in $\Theta_2[\{a^{\beta{,}0}_i\}]$ arising from the incompatibility between the pre- and post-quench Hamiltonians, as discussed in Sec.~\ref{sub:fdr}. 
Therefore, it is essential that the homogeneous system satisfies the condition: $[\hat{\mathcal{H}}_{0}, \sum_{i=1}^N \hat{A}_i] \neq 0$. 
Otherwise, $\Theta^{\msc{hom}}_2[a^{\beta{\,}0}]$ would be identically zero.

\subsection{Approximation for $\Theta_2[\{a^{\beta{\,}0}_i\}]$ via perturbation treatment of the KS system}\label{aprxks}
The fictitious KS world is governed by a Hamiltonian that differs from the one describing the actual system under investigation.
However, if the exact form of the functional  ${\lambda}^{\msc{h-xc}}_i[\{a^\beta_i\}]$ is known, the KS framework can accurately reproduce the density of the original interacting system.  
Therefore, a well-founded idea is to treat the KS Hamiltonian as a zeroth-order approximation to the `\textit{true}' Hamiltonian~\cite{herrera2018melting,herrera2017dft,gorling1994exact,gorling1993correlation}. 
This is because the KS Hamiltonian encapsulates key aspects of the many-body properties inherent the real system.
In this perturbation-like approach, we can express the original Hamiltonian as $\hat{\mathcal{H}} = \hat{{\mathcal{H}}}^{\msc{ks}} + \Delta\hat{\mathcal{H}}$, where $\Delta\hat{\mathcal{H}} = \hat{\mathcal{H}} - \hat{{\mathcal{H}}}^{\msc{ks}}$.
Accordingly, any observable property $\mathcal{Q}[\{a^{\beta{\,}0}_i\}]$ can be expanded as
\begin{equation}
\mathcal{Q}[\{a^{\beta{\,}0}_i\}] = {\mathcal{Q}}^{\msc{ks}}[\{a^{\beta{\,}0}_i\}] + \Delta \mathcal{Q}[\{a^{\beta{\,}0}_i\}],
\end{equation}
where ${\mathcal{Q}}^{\msc{ks}}[\{a^{\beta{\,}0}_i\}]$ represents the zeroth-order approximation of the quantity of interest. 
This method is particularly advantageous when the LDA scheme is not applicable. 
For example, if the underlying homogeneous system satisfies  $[\hat{\mathcal{H}}_{0},\sum^N_{i=1}\hat{A}_i]= 0$, the zeroth-order approximation can effectively address the incompatibility between the initial and final forms of the interacting Hamiltonian.
In such cases, $\Theta_2[\{a^{\beta{\,}0}_i\}]$ can be approximated using the corresponding quantity calculated within the KS framework.

\section{The Kohn-Sham scheme in the canonical ensemble} \label{sec5}
In Sec.~\ref{sec4}, we introduced a finite-temperature KS mapping, which enables the accurate evaluation of equilibrium thermal densities through iterative solutions of the self-consistent MKS equations, i.e., Eqs.(\ref{KS1}) and~(\ref{KS}). 
However, dealing with statistical systems that have a fixed number of indistinguishable particles, even within the non-interacting KS framework, presents considerable challenges~\cite{PhysRevLett.84.4255,borrmann1993recursion,PhysRevE.83.067701,PhysRevResearch.2.043206}, making the solution of the MKS equations computationally demanding.
This complexity is one reason why finite-temperature DFT calculations typically employ the grand canonical ensemble, where the average number of particles is fixed by $\braket{N}=\sum_i f_\mu(\epsilon_i)$. 
In a many-fermion system, the Fermi-Dirac distribution $f_\mu(\epsilon_i)=(1+e^{\beta(\epsilon_i-\mu)})^{-1}$ describes the occupation probabilities of the KS eigenstates, with $\epsilon_i$ representing the KS eigenvalues and $\mu$ the chemical potential.  
This framework offers the advantage of a straightforward expression for thermal electron densities: 
$n^\beta(\mathbf{r})=\sum_i f_\mu(\epsilon_i)|\psi^{\msc{ks}}_i(\mathbf{r})|^2$, based on the KS wavefunctions $\psi^{\msc{ks}}_i(\mathbf{r})$~\cite{pribram2014thermal}. 
Here, we present a method to compute thermal densities within the canonical ensemble using the KS mapping while keeping the calculations feasible.  

To elucidate our approach, we consider a system of $N$ interacting fermions on a lattice, characterized by the Hamiltonian~(\ref{Hl}), with the specific form
\begin{equation}
\hat{\mathcal{H}}=\hat{\mathcal{H}}_0+\sum^{\mathcal{L}}_{i=1} V_i \hat{n}_i.
\label{HVn}
\end{equation}
The universal part of this Hamiltonian reads
\begin{equation}
\hat{\mathcal{H}}_0=\hat{\mathcal{T}}+\hat{\mathcal{W}},
\label{hh}  
\end{equation}
where
\begin{equation}
\hat{\mathcal{T}}=-J
\sum_{i; \sigma=\uparrow,\downarrow}(\hat{c}^\dagger_{i,\sigma}\hat{c}_{i+1,\sigma} + \mathrm{h.c.})
\label{tt}  
\end{equation}
is the kinetic term, while $\hat{\mathcal{W}}$ accounts for the two-body interaction.
In Eq.~(\ref{tt}), $\hat{c}^\dagger_{i,\sigma}$ and $\hat{c}_{i,\sigma}$ denote the creation and annihilation operators for a fermion with spin $\sigma=\uparrow,\downarrow$, and $\hat{n}_i=\hat{n}_{i,\uparrow}+ \hat{n}_{i,\downarrow}$ is the total number operator for the $i$-th site.

Given the form of the external potential, the thermal densities are naturally defined as  $n^\beta_i = \Tr\{\hat{\rho}^{\msc{th}}_\beta[\{V_i\}]\hat{n}_i\}$. 
According to the MHK theorem, the following correspondence holds:
\begin{equation}
 \{V_i\} \iff \{n^\beta_i\} \iff \hat{\rho}^{\msc{th}}_\beta \equiv \hat{\rho}^{\msc{th}}_\beta[\{n^\beta_i\}].
\end{equation}
As remarked in Sec.~\ref{sec4}, there exists a KS system having exactly the same set of thermal densities as the original interacting system. 
The corresponding KS Hamiltonian is
\begin{equation}
\hat{\mathcal{H}}^{\msc{ks}}=\hat{\mathcal{T}}+\sum^{\mathcal{L}}_{i=1}({V}^{\msc{h-xc}}_i[\{n^\beta_i\}]+V_i)\hat{n}_i.
\label{KKS}
\end{equation}
At this point, we seek a more computationally tractable equation for the thermal densities than Eq.~(\ref{KS}). 
In particular, we build on previous research~\cite{PhysRevLett.84.4255,borrmann1993recursion} that examined the canonical partition function for $N$ non-interacting fermions.
These studies enable us to express the canonical partition function for the KS fermions using the following recursive formula: 
\begin{equation}
\mathcal{Z}_N = \frac{1}{N}\sum^{N}_{m=0}(-1)^{m-1}\mathcal{Z}_1(m\beta)\mathcal{Z}_{N-m}(\beta), 
\end{equation}
where $\mathcal{Z}_1(0)=1$ and $\mathcal{Z}_1(m\beta)=\sum_i e^{-m\beta\epsilon_i}$ for $m>1$.

Given the structure of the KS Hamiltonian~(\ref{KKS}), the partition function for a KS system with $N_\uparrow$ spin-up and $N_\downarrow$ spin-down fermions becomes:
$\mathcal{Z}^{\msc{ks}}_N=\mathcal{Z}^{\msc{ks}}_{N_\uparrow}\mathcal{Z}^{\msc{ks}}_{N_\downarrow}$, 
with particle number conservation ensured by $N = N_\uparrow + N_\downarrow$.
The corresponding equilibrium free energy is then: 
$\mathcal{F}^{\msc{ks}}=-\frac{1}{\beta}\log(\mathcal{Z}^{\msc{ks}}_N)$, 
from which the equilibrium thermal densities can be extracted using the HF theorem as
\begin{equation}
n_i^\beta= \frac{\partial \mathcal{F}^{\msc{ks}}}{\partial V^{\msc{ks}}_i}.
\label{KS3}
\end{equation}
The combination of Eq.~(\ref{KS3}) and Eq.~(\ref{KS1}) forms the self-consistent foundation of our finite-temperature KS approach, which, in principle, exactly reproduces the thermal densities of the original interacting system.

\section{A notable example\label{sec6}}
This section is dedicated to validating our finite-temperature KS approach, as defined by Eqs.~(\ref{KS1}) and~(\ref{KS3}), within the context of the Hubbard model~\cite{gutzwiller1963effect,hubbard1963electron,kanamori1963electron}.
Specifically, we examine a scenario where electrons are influenced by an inhomogeneous external potential dependent on the parameter $\msc{v}_0$.
The system Hamiltonian is expressed as: 
\begin{equation}
\hat{\mathcal{H}}=\hat{\mathcal{H}}_0+\sum^{\mathcal{L}}_{i=1} V_i(\msc{v}_0) \hat{n}_i,
\label{dim}
\end{equation}
where $V_i(\msc{v}_0) = f_i \msc{v}_0$ with the dimensionless set $\{f_i\}$ that defines the spatial shape of the external potential along the chain. Here  $\hat{\mathcal{H}}_0$ takes the form given in Eq.~(\ref{hh}), with the two-body interaction term given by:  
\begin{equation}
\hat{\mathcal{W}}=\sum^{\mathcal{L}}_{i=1} U\hat{n}_{i,\uparrow}\hat{n}_{i,\downarrow}.
\end{equation}
As described in Sec.~\ref{sec3} the system is  initially prepared in the Gibbs state  $\hat{\rho}^{\msc{th}}_\beta[\{n^\beta_i\}]$.  
Subsequently, it is decoupled from the thermal bath and undergoes an instantaneous quench in the work parameter, with amplitudes $\delta V_i = f_i \delta \msc{v}_0$.

We focus on the half-filled Hubbard model, with the total spin along the $z$ direction set to zero. 
Our analysis considers two external potentials. One that decreases linearly along the chain, 
\begin{equation}
\hat{\mathcal{V}}_{\msc{ext}} = \sum^{\mathcal{L}}_{i=1}\left[\msc{v}_0 - \frac{2\msc{v}_0(i-1)}{\mathcal{L}-1}\right] \hat{n}_i, 
\label{ext}
\end{equation}
and another one with a harmonic dependence,  
\begin{equation}
\hat{\mathcal{V}}_{\msc{ext}} = \sum^{\mathcal{L}}_{i=1} \frac{1}{2}\msc{v}_0\left[i - \frac{\mathcal{L}+1}{2}\right]^2\hat{n}_i. 
\label{ext1}
\end{equation} 
We begin by analyzing the exact results for the Hubbard dimer, as presented in Sec.~\ref{subH2}. We then compare these results with those obtained from our KS  mapping, as detailed in Sec.~\ref{subH2KS} and further explored in Sec.~\ref{KShubdim}.
Next, we extend our study to longer Hubbard chains probed by the linear potential defined in Eq.~(\ref{ext}), as outlined in Sec.~\ref{subH2KS} and Sec.~\ref{subExW}. 
In particular, in Sec.~\ref{subH2KS}, we compare exact results for systems with up to $8$ sites with corresponding ones from our KS mapping. 
Finally, in Secs.~\ref{subExWLP} and \ref{subExWHP}, we investigate the impact of electron-electron interactions on work extraction in longer chains, considering both linear and harmonic potentials given in Eqs.~(\ref{ext}) and~(\ref{ext1}).

\subsection{Exact results for the Hubbard dimer\label{subH2}}
In the two-particle subspace with total spin zero along the $z$-axis, the Hamiltonian~(\ref{dim}) characterizes a two-site Hubbard chain and is represented by the matrix
\begin{equation}
\hat{\mathcal{H}} \dot{=} 
\begin{pmatrix}
U+2V_1 & -J & J & 0\\
-J & V_1+V_2 &0  &-J\\
J& 0 & V_1+V_2 &J\\
0& -J & J & U+2V_2 
\end{pmatrix},
\end{equation}
in the basis $\{\ket{\uparrow \downarrow, 0}, \ket{\uparrow, \downarrow}, \ket{\downarrow, \uparrow}, \ket{0, \uparrow \downarrow}\}$.
This straightforward, exactly solvable model exhibits a diverse range of physical phenomena~\cite{herrera2017dft,PhysRevLett.114.080402,editorial2013hubbard,carrascal2015hubbard}, including a precursor to the Mott metal-insulator transition and, influenced by the external potential, a precursor to the ionic insulator transition.
The two transitions are in competition, with the former favoring single-site occupation and the latter promoting double-site occupation.
The metal phase emerges in the narrow region where $U \sim 2\msc{v}_0$, driven by the interplay of the interaction term and the external potential.

\begin{figure}[!h]
\centering
 \begin{minipage}[b]{0.49\textwidth}
 \centering
 \includegraphics[width=\textwidth]{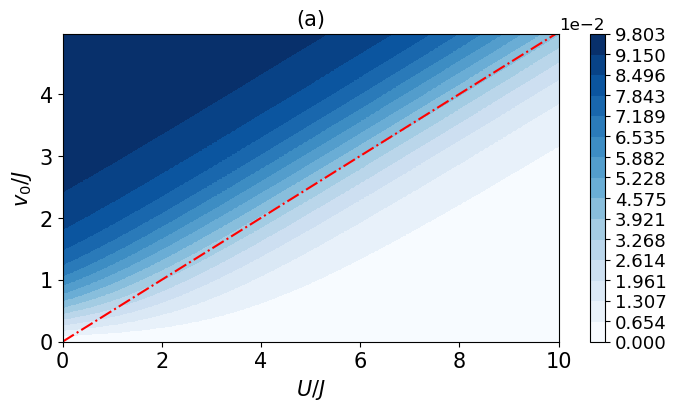}
 \end{minipage}
 \begin{minipage}[b]{0.49\textwidth}
 \centering
 \includegraphics[width=\textwidth]{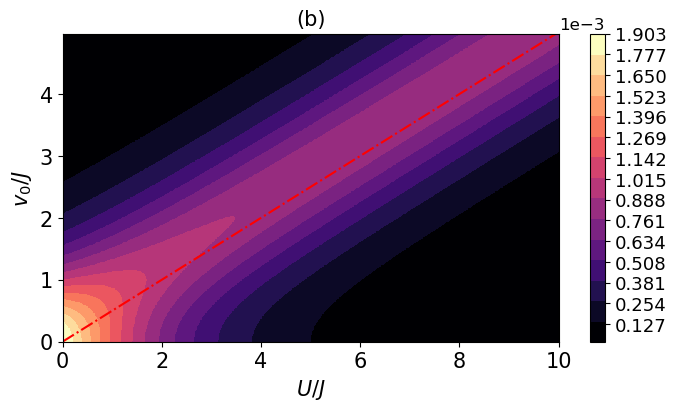}
 \end{minipage}
 \begin{minipage}[b]{0.49\textwidth}
 \centering
 \includegraphics[width=\textwidth]{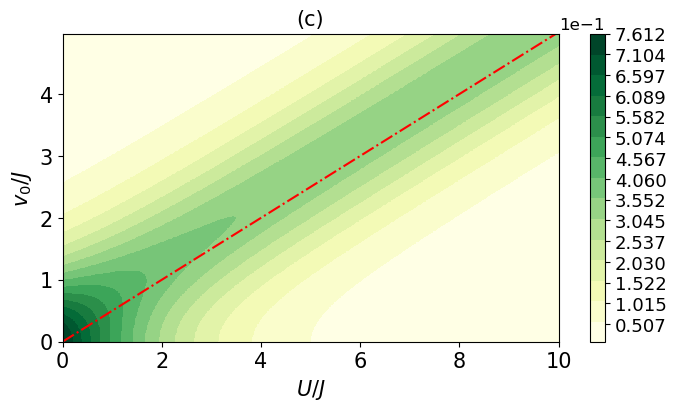}
\end{minipage}
\caption{(a) Average extracted quantum work, $\braket{w}_{\msc{ex}}$, (b) average irreversible entropy production, $\braket{\mathcal{S}_{\msc{irr}}}$, and~(c) the first derivative of the thermal density $n^\beta_i$ with respect to the work parameter $\msc{v}_0$, for a Hubbard dimer at the inverse temperature $\beta = 1/J$ and a sudden quench of amplitude $\delta \msc{v}_0 = 0.05J$. 
In all panels, the red dashed-dotted line represents the condition $U = 2\msc{v}_0$.
\label{fig1}}
\end{figure}

In Fig.~\ref{fig1}(a), we examine the average extracted work, $\braket{w}_{\msc{ex}} = -\braket{w}$, as a function of $U$ and $\msc{v}_0$, following a sudden quench of amplitude $\delta v = 0.05J$. 
For $U > 2\msc{v}_0$, the system enters the Mott insulating phase, leading to a decrease in the extractable work as the interaction strength $U$ increases.
In this phase, double occupancy of sites becomes energetically unfavorable, rendering work extraction through the external potential quench impractical. 
This behavior can be understood by examining the thermal densities in the limit of large $U$: for $U \gg 2\msc{v}_0$, we have $n^\beta_1 \sim n^\beta_2 \sim 1$, which, according to Eq.(\ref{ws}), results in $\braket{w} \sim 0$.
Conversely, for $U < 2\msc{v}_0$, the system is in the ionic insulating phase, where the extractable work increases as $U$ decreases, reaching its maximum value as $U$ approaches zero.
This is because for $U\ll2\msc{v}_0$ we find  $n^\beta_1\sim 2$ and $n^\beta_2\sim 0$, leading to $\braket{w} \sim -2\delta \msc{v}_0$ as per Eq.~(\ref{ws}).

In Fig.~\ref{fig1}(b), we show the average  irreversible entropy production $\braket{\mathcal{S}_{\msc{irr}}}$ for the same process.
As expected, $\braket{\mathcal{S}_{\msc{irr}}}$  exhibits a pronounced peak in the metallic region separating the Mott insulating phase from the ionic insulating phase. 
This behavior can be understood through the dependence of the average irreversible entropy production on the thermal densities, as described in Eq.~(\ref{Sis}). 
The thermal densities are sensitive to small variations in the work parameter $\msc{v}_0$ when $U \sim 2\msc{v}_0$, which is also reflected in the peak observed in the derivatives of the thermal density with respect to $\msc{v}_0$, shown in Fig.~\ref{fig1}(c).

\subsection{Thermal KS mapping for Hubbard chains\label{subH2KS}}

\begin{figure*}[htbp]
    \centering
    \includegraphics[width=0.9\textwidth]{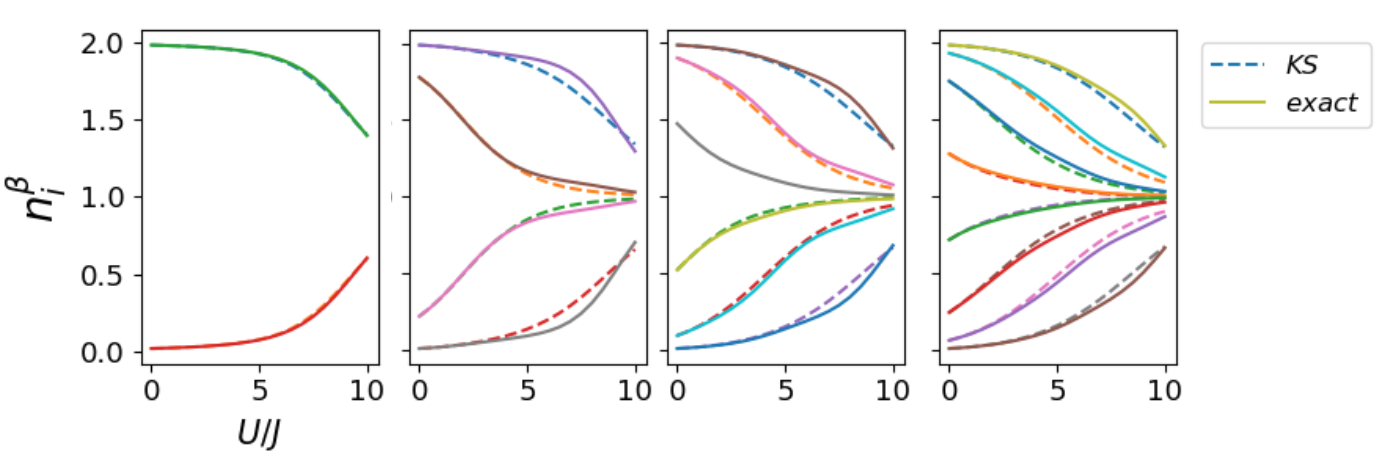}
    \caption{Thermal densities obtained through exact diagonalization of the Hubbard Hamiltonian~(\ref{dim}) compared with those from our KS mapping, as a function of $U$ for $\msc{v}_0 = 5J$ and $\beta J = 1$. The percentage error for chains of $2$, $4$, $6$, and $8$ sites is $0.7\%$, $3.7\%$, $2.28\%$, and $1.78\%$, respectively.}
    \label{figd}
\end{figure*}

To evaluate the thermodynamic quantities of interest for the Hubbard model using our thermal KS scheme, the first step is to construct an accurate approximation for the H-XC potential. 
Some previous studies~\cite{SPB16, carrascal2015hubbard, PhysRevA.92.013614, PhysRevB.86.235139} provide a robust framework for this construction. 
Furthermore, in the context of the Anderson model for a single nonmagnetic impurity coupled to two leads, it has been demonstrated that the impurity Hamiltonian, modeled as a single-site Hubbard model, is $v$-representable in the non-interacting case. This allows for the derivation of an analytical form for the XC potential~\cite{PhysRevLett.107.216401}.

To proceed further, we treat the Hubbard chain as a collection of single sites coupled by the kinetic term~(\ref{tt}). Following the LDA, we approximate the KS potential as:
\begin{align}
{V}^{\msc{ks}}_i[\{n^\beta_i\}]\simeq & V_i+ {V}^{\msc{h-xc}}_{\msc{ss}}[n^\beta_i]= \notag\\=& V_i+ U+\frac{1}{\beta}\ln{\Gamma^{\beta}_{U}[n^\beta_i]},
\label{ldahub}
\end{align}
where ${V}^{\msc{h-xc}}_{\msc{ss}}[n^\beta]$ is the single-site H-XC potential, derived from the nonmagnetic impurity Anderson model~\cite{PhysRevLett.107.216401}. 
Further details of this construction are provided in appendix \ref{app4}, with the explicit form of $\Gamma^{\beta}_{U}$ given in Eq.~(\ref{GambU}).
It is worth mentioning that Eq.~(\ref{ldahub}) can be viewed as a simplified version of the LDA scheme implemented in 
\cite{PhysRevB.86.235139}, which utilized the solution of the homogeneous problem via the thermal Bethe ansatz.

With a reliable approximation for the H-XC potential in hand, we can solve our thermal KS equations: 
\begin{align}
&\{\hat{\mathcal{T}}+\sum_i{V}^{\msc{ks}}[n^\beta_i]\hat{n}_i\}\ket{\phi^{i}_{\beta}}=\epsilon^{i}_{\beta}\ket{\phi^{i}_{\beta}} 
\label{hubKS1}\\
&n^\beta_i=-\frac{2}{\beta}\frac{\partial \log(\mathcal{Z}^{\msc{ks}}_{N_\uparrow})}{\partial V^{\msc{ks}}_i},
\label{hubKS2}
\end{align}
adapted from Eqs.~(\ref{KS}) and~(\ref{KS3}), to obtain the set of thermal densities.

Figure \ref{figd} illustrates a comparison between the thermal densities obtained through exact diagonalization of Eq.~(\ref{dim}) and those obtained using our KS mapping, as a function of the interaction parameter $U$.
This analysis covers chains with up to $8$ sites, using the external potential defined in Eq.(\ref{ext}), with $\msc{v}_0 = 5J$ and the inverse temperature $\beta = 1/J$.
To assess the accuracy of the approximation, we use the density metric discussed in~\cite{PhysRevLett.106.050401}
\begin{equation}
\mathcal{D}(\{n^\beta_i\},\{n^{\beta{\msc{ks}}}_i\})= \frac{1}{2N}\sum_i |n^\beta_i-n^{\beta{\msc{ks}}}_i|,
\end{equation}
where the scaling factor $2N$ represents the maximum possible distance between the two sets of thermal densities, ensuring that the distances lie within the range $[0,1]$.
For the chosen density metric and parameters, the percentage error for all the chain lengths is less than $3.7\%$.

Now that the validity of the approximation for  the thermal densities is established, we can proceed to evaluate the thermodynamic quantities of interest for which an explicit functional is available.

\subsection{Thermal KS mapping for the Hubbard dimer\label{KShubdim}}
Here we leverage our KS mapping, defined by the above derived self-consistent equations~(\ref{hubKS1}) and~(\ref{hubKS2}), to compute various statistical moments for the Hubbard dimer and compare these with the exact solution discussed in Sec.~\ref{subH2}. 
Specifically, we focus on evaluating the first and second moments of the PDW and the first moment of the PDE.

\begin{figure}[h!]
    \centering
    \begin{minipage}[b]{0.49\textwidth}
        \centering
        \includegraphics[width=\textwidth]{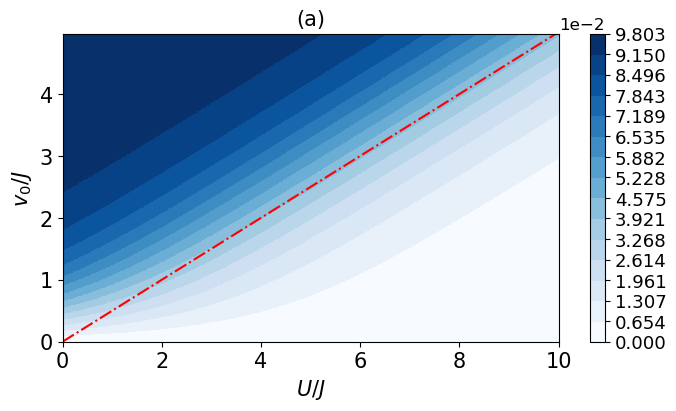}
    \end{minipage} 
    \centering
    \begin{minipage}[b]{0.49\textwidth}
        \centering
        \includegraphics[width=\textwidth]{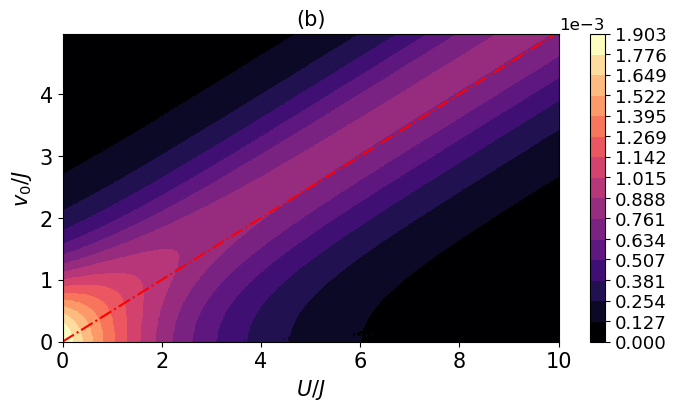}
    \end{minipage} 
    \caption{Mean values of (a)~the extracted quantum work $\braket{w}_{\msc{ex}}$ and (b)~the irreversible entropy production  $\braket{\mathcal{S}_{\msc{irr}}}$, for a Hubbard dimer at $\beta=1/J$ and $\delta \msc{v}_0=0.05J$. In both panels, the red dashed line corresponds to $U = 2\msc{v}_0$.}
    \label{fig2}
\end{figure}

\begin{figure}[!h]
    \centering
    \begin{minipage}[b]{0.45\textwidth}
        \centering
        \includegraphics[width=\textwidth]{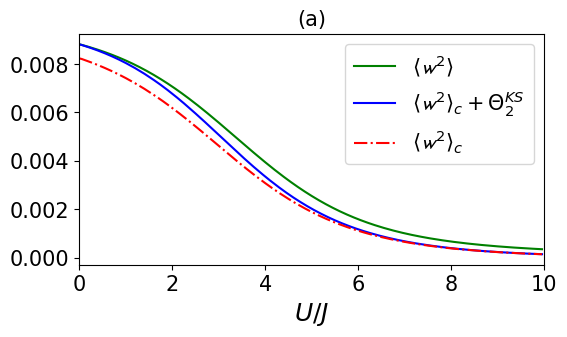}
    \end{minipage}
 \centering
    \begin{minipage}[b]{0.45\textwidth}
        \centering
        \includegraphics[width=\textwidth]{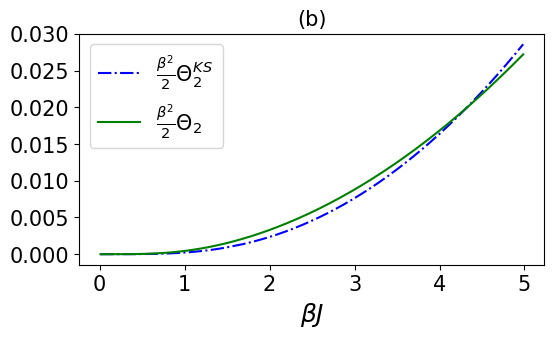}
    \end{minipage}
    \caption{(a)~Second moment of the PDW for the Hubbard dimer, $\braket{w^2}$, as a function of the interaction strength $U$, at the inverse temperature of $\beta = 1/J$. 
    This exact result is compared with the KS result  $\braket{w^2}_{c} + \Theta^{\msc{ks}}_2$ and the contribution $\braket{w^2}_{c}$ alone. 
    (b)~Non-commutative functional $\Theta_2$ evaluated exactly and via our KS scheme as a function of the inverse temperature $\beta$, for $U=3J$. 
    In both panels, the sudden quench protocol $\msc{v}_0 = 2.00J \rightarrow v_f = 2.05J$ is considered, corresponding to a sudden quench of amplitude $\delta \msc{v}_0 = 0.05J$.
    \label{fig3}}
\end{figure}

We determine the average extractable work, $\braket{w}_{\msc{ex}} = -\braket{w}$, using Eq.~(\ref{ws}), and the average irreversible entropy production, $\braket{\mathcal{S}_{\msc{irr}}}$, using Eq.~(\ref{Sis}), from the thermal densities obtained by solving Eqs.~(\ref{hubKS1}) and~(\ref{hubKS2}).
Figure~\ref{fig2}(a) and~\ref{fig2}(b) display $\braket{w}_{\msc{ex}}$ and $\braket{\mathcal{S}_{\msc{irr}}}$, respectively,  as functions of the interaction strength $U$ and the work parameter $\msc{v}_0$. 
The system is set at an inverse temperature of $\beta = 1/J$ with a sudden quench of amplitude  $\delta \msc{v}_0 = 0.05J$.
The computed results are in excellent agreement with the exact results shown in Fig.~\ref{fig1}(a) and~\ref{fig1}(b), validating the reliability and accuracy of our KS framework for the Hubbard dimer.
This approach becomes comparatively less precise when computing the second moment of the PDW.
As outlined in Sec.~\ref{sec3}, $\braket{w^2}$ consists of two contributions: $\braket{w^2}_c$, which is an explicit functional of the thermal density, as defined in Eq.~(\ref{w^2}), and $\Theta_2$, introduced in Eqs.~(\ref{w^22}) and~(\ref{FDR2}), which addresses the incompatibility of the pre-quench and post-quench Hamiltonians. 
Currently, there is no available explicit functional form for $\Theta_2$.  
Besides, the Hubbard model falls into the category where an LDA approach for this contribution cannot be implemented because $[\hat{\mathcal{H}}_{0},\sum^N_{i=1}\hat{n}_i]= 0$. 
Thus, we use the zeroth-order approximation  described in Sec.~\ref{aprxks}, and apply the generalized FDR of Eq.~(\ref{FDR2}) to estimate the non-commutative functional as: ${\Theta}_2[\{n^{\beta,0}_i\}]\approx{\Theta}^{\msc{ks}}_2[\{n^{\beta,0}_i\}]$. 
Figure~\ref{fig3}(a) shows $\braket{w^2}$, as calculated within the analytic framework of the Hubbard dimer (green line). 
This exact result is compared with $\braket{w^2}_{\msc{ks}}=\braket{w^2}_{c}+{\Theta}^{\msc{ks}}_2$, obtained via our KS mapping (blue line), and with $\braket{w^2}_{c}$ alone (red line).
In particular, $\braket{w^2}$, $\braket{w^2}_{\msc{ks}}$ and $\braket{w^2}_{c}$ are plotted as a function of $U$ at an intermediate temperature of $\beta J = 1$ and a sudden quench of amplitude $\delta \msc{v}_0 = 0.05J$. 
We observe that our KS approach becomes less accurate as the value of $U$ increases above $\sim 2J$. 
In fact, for $U\gtrsim 5$  $\braket{w^2}_{\msc{ks}}$ appears indistinguishable from $\braket{w^2}_{c}$, and the contribution from the incompatibility between the pre- and post-quench KS Hamiltonians vanishes.
The underlying reason is that, as the parameter $U$ increases, the thermal densities $n_1^\beta$ and $n_2^\beta$ tend to be equal, making ${V}^{\msc{ks}}[n^\beta_i] \approx {V}^{\msc{h-xc}}_{\msc{ss}}[n^\beta_i]$, and thus $[\hat{\mathcal{H}}, \sum_i {V}^{\msc{ks}}[n^\beta_i] \hat{n}_i] = 0$.

To assess the accuracy of our KS scheme as a function of temperature, Fig.~\ref{fig3}(b) displays ${\Theta}^{\msc{ks}}_2[\{n^{\beta,0}_i\}]$ and ${\Theta}_2[\{n^{\beta,0}_i\}]$ as functions of $\beta J$ for $U = 3J$. 
The two results exhibit values that are close to each other and show similar trends, indicating that the approximation remains reliable across the explored temperature range for the specified value of $U$.

\subsection{Average extracted work for Hubbard chains with varying length\label{subExW}}

\begin{figure*}[htbp]
    \centering
    \includegraphics[width=0.9\textwidth]{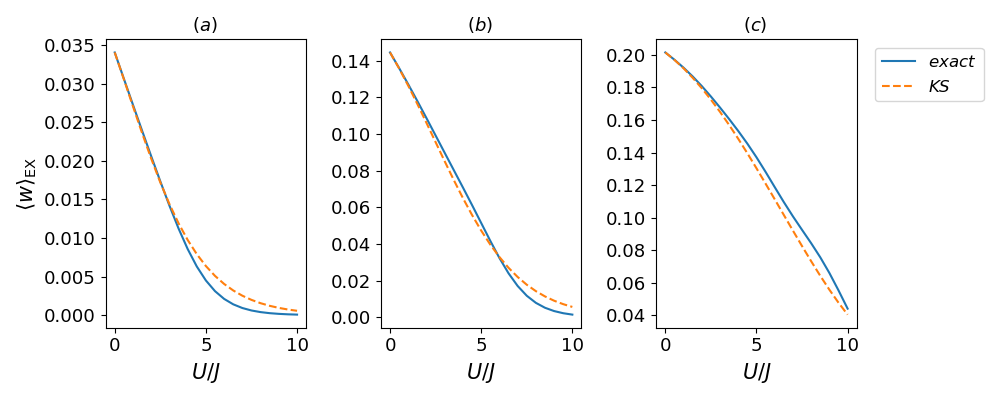}
    \caption{Average work extracted with the linear potential defined in Eq.~(\ref{ext}) for a sudden quench of amplitude $\delta \msc{v}_0 = 0.05J$. 
    Exact results obtained via diagonalization of the Hamiltonian~(\ref{dim}) are compared with those from our KS mapping for a chain with $8$ sites. 
    The comparisons are made for three different values of $\msc{v}_0$, namely: (a)~$\msc{v}_0 = 0.5J$, (b)~$\msc{v}_0 = 2.5J$, and (c)~$\msc{v}_0 = 5J$. 
    The corresponding maximum errors are: (a)~$0.0019J$, (b)~$0.0063J$, and (c)~$0.0113J$.}
    \label{fig6}
\end{figure*}

\begin{figure}[htbp]
    \centering
    \begin{minipage}[b]{0.49\textwidth}
        \centering
        \includegraphics[width=\textwidth]{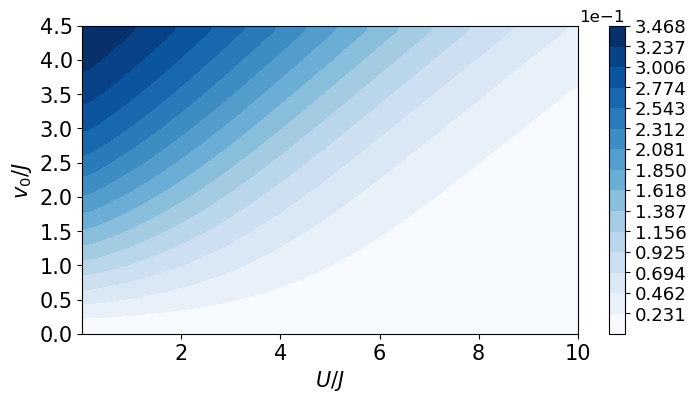}
    \end{minipage} 
    \caption{Average work extracted from a Hubbard chain with $16$ sites, probed by the linear potential of Eq.~(\ref{ext}). 
    The system is at an inverse temperature of $\beta = 1/J$, with a sudden quench of amplitude $\delta \msc{v}_0 = 0.05J$.}
    \label{fig7}
\end{figure}
We finally focus  on the possibility of extracting work from an interacting many-body system.   
According to Eq.~(\ref{ws}) the average work for the Hubbard chain is given by
\begin{equation}
\braket{w}=\sum_i f_i\delta \msc{v}_0 n^\beta_i
\end{equation}  
where the sign of the quenching amplitude $\delta \msc{v}_0$ and the behavior of the thermal densities $n^\beta_i$ determine whether the work can be extracted, meaning $\braket{w} < 0$. 
In this context, the thermal densities are strongly influenced by the interplay between the external potential, characterized by the set of parameters $\{f_i \msc{v}_0\}$, and the many-body interaction.
As observed for the Hubbard dimer in Sec.~\ref{subH2}, the system can experience different phases depending on the relative magnitudes of $U$ and $\msc{v}_0$. 
For $\msc{v}_0 \gg U$, a band insulator phase emerges, where double occupancy is favored. 
For $U \gg \msc{v}_0$, a Mott insulator phase occurs, where the strong interaction inhibits double occupancy. 
When $U$ and $\msc{v}_0$ are comparable, a metallic phase manifests, due to the competition between the external potential and the many-body interaction.
In this respect, the external potential plays a crucial role in determining which phase occurs in different regions of the chain.
We here demonstrate that the external potential can actually enhance the effects of many-body interactions, thereby facilitating the extraction of work, either by driving the system towards the band insulator phase or by shifting it towards the Mott insulator phase.
As outlined earlier in this section, we focus our analysis on the linear decreasing potential~(\ref{ext}) and the harmonic potential~(\ref{ext1}), establishing that the ability to extract work from many-body interactions is highly dependent on the nature of the external potential.

\subsubsection{Linear Potential} \label{subExWLP}
We begin by considering the external potential that decreases linearly along the chain. 
In Fig.~\ref{fig6}, we compare the extracted work, calculated exactly and using our KS mapping for an $8$-site chain. 
As expected, given the good approximation for the thermal densities, discussed in Fig.~\ref{figd}, the two results show excellent agreement. 
Figure~\ref{fig6} also illustrates that for a fixed value of $\msc{v}_0$, and a quench of amplitude $\delta \msc{v}_0 > 0$, the extractable work decreases monotonically as the interaction strength $U$ increases. 

This behavior is further explored in the context of a $16$-site chain.
In Fig.~\ref{fig7}, the extracted work is shown as a function of $U$ and $\msc{v}_0$. 
With the system prepared in the band insulator phase, near-double occupancy is favored at the sites in the second half of the chain, although this is partially masked by finite temperature effects. 
Performing a quench of the work parameter with an amplitude $\delta \msc{v}_0 >0$ further promotes this double occupancy, which translates into the opportunity of extracting the maximum possible work: 
\begin{equation}
\braket{w} \sim 2\delta \msc{v}_0 \sum^{\mathcal{L}}_{i=\mathcal{L}/2}
\left[1-\frac{2(i-1)}{\mathcal{L}-1}\right] = -\frac{(\mathcal{L}^2-4)\delta \msc{v}_0}{2(\mathcal{L}-1)}. 
\end{equation}

Conversely, at a fixed initial $\msc{v}_0$, as the interaction parameter increases, less and less work can be extracted from the system. 
This occurs because the electron-electron interaction opposes double occupancy, counteracting the effect of the quench $\msc{v}_0 \rightarrow \msc{v}_0+\delta \msc{v}_0 $ that would otherwise favor it. 
In the strongly interacting limit, where each $n^\beta_i \sim 1$, the external potential form prevents work from being extracted at all:
\begin{equation}
\braket{w} \sim \delta \msc{v}_0 \sum^{\mathcal{L}}_{i=1}\left[1-\frac{2(i-1)}{\mathcal{L}-1}\right] =0. 
\end{equation}

In summary, for a positive quench of amplitude $\delta \msc{v}_0 > 0$, the work lies in the range $-\frac{(\mathcal{L}^2-4)\delta \msc{v}_0}{2(\mathcal{L}-1)} \leq \braket{w} \leq 0$.  
This indicates that work can be extracted starting from any initial state, with many-body interactions generally impeding this extraction.
On the other hand, for quenches of negative amplitudes, $\delta \msc{v}_0 < 0$, the  work is positive covering  the range $0 \leq \braket{w} \leq \frac{(\mathcal{L}^2-4) |\delta \msc{v}_0|}{2(\mathcal{L}-1)}$.

\subsubsection{Harmonic potential} \label{subExWHP}
\begin{figure}[htbp]
    \centering
    \begin{minipage}[b]{0.35\textwidth}
        \centering
        \includegraphics[width=\textwidth]{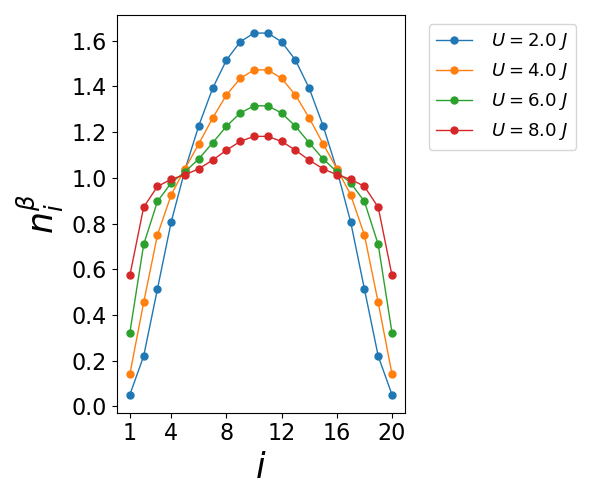}
    \end{minipage} 

    \caption{Spatial distribution of the thermal densities for a fixed $\msc{v}_0 = 0.175J$ and four different values of $U$ at an inverse temperature $\beta J = 1$.}
    \label{fig9}
\end{figure}

We now turn our attention to the scenario where electrons are subjected to a parabolic~(harmonic) potential.   
The impact of this potential on the thermal densities $\{n_i^\beta\}$ is illustrated in Fig.~\ref{fig9} for a fixed $\msc{v}_0$ and four distinct values of $U$ on a $20$-site chain.
It is evident that the primary impact of many-body interactions is to suppress the double occupancy at the center of the chain favored  by the external potential.

Unlike the linear potential case, the parabolic potential does not allow work extraction via a quench that drives the system toward a band insulator phase, regardless of the values of $\msc{v}_0$ and $U$, i.e., independent of the phase in which the system resides before the quench. 
In other words, no extraction of work is possible for a quench in which the parabola amplitude is changed from $\msc{v}_0$ to $\msc{v}_0+\delta \msc{v}_0$.

\begin{figure}[htbp]
    \centering
    \begin{minipage}[b]{0.50\textwidth}
        \centering
        \includegraphics[width=\textwidth]{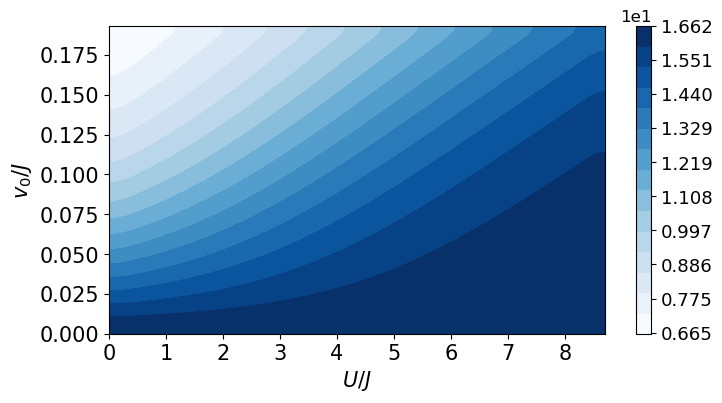}
    \end{minipage} 
    \caption{Mean value of the quantum work extracted from a Hubbard chain subjected to a harmonic potential, at an inverse temperature $\beta = 1/J$, following a quench with amplitude $\delta \msc{v}_0 = 0.05J$.}
    \label{fig8}
\end{figure}
Conversely, if the quench increases the parabola  amplitude, the average work $\braket{w}$ is negative for any initial values of $\msc{v}_0$ and $U$, meaning that work can always be extracted.
The underlying reason is that the Coulomb repulsion and the widening of the parabola do not compete but rather favor the condition $\{n_i^\beta \sim 1 \}$. 
In contrast to the linear potential, stronger many-body interactions here result in more work being extracted. Specifically, in the deep Mott phase, the extractable work reaches its maximum possible value.

\section{Conclusions}
\label{sec7}
In this work, we have explored the thermodynamics of many-body systems through the lens of the MHK theorem and a finite temperature KS mapping.
In particular, utilizing the MHK theorem, we demonstrated that the PDW and the PDE, along with their characteristic functions, can be expressed as functionals of the thermal densities.
This insight turned to be particularly powerful as it provides a unified framework for analyzing the statistics of work and irreversible entropy production in the context of sudden quenches of small amplitudes.
Specifically, we derived simple analytical expressions for the first moments of the PDW and the PDE that are functionals of the thermal densities and their derivatives. 
For the second moment of the PDW, we identified and separated two distinct contributions: $\braket{w^2}_c$, a classical-like term, and $\Theta_2[{n^{\beta,0}_i}]$, a purely quantum contribution arising from the non-commutativity of the pre- and post-quench Hamiltonians. 
This separation enabled us to establish a generalized FDR in the sudden quench regime to the lowest order in the amplitude parameters ${\delta\lambda_i}$.
We then introduced a KS mapping as a promising approach for studying the thermodynamics of many-body systems, especially when explicit functionals for thermodynamic quantities are unavailable. 
In Section~\ref{sec5}, we developed a method to evaluate thermal densities using the KS scheme within the canonical ensemble framework, thereby bypassing the need to construct the many-body Gibbs state explicitly.
The proposed method was validated on out-of-equilibrium thermodynamics of the Hubbard model under the influence of a non-homogeneous external potential. 
We assessed the accuracy of the KS mapping by comparing the exact thermal densities with those obtained via the KS approach for chains of up to $8$ sites. 
Furthermore, we explored the quantum thermodynamics of the quenched Hubbard dimer, illustrating the applicability and limitations of the KS mapping when an explicit functional is unavailable.
Our investigation into the role of many-body interactions in work extraction revealed that the KS mapping, combined with our general findings, provides a robust framework for analyzing the thermodynamic properties of complex quantum systems. 
In summary, our KS mapping, along with the insights gained from the MHK theorem, offers a powerful tool for studying the thermodynamics of many-body systems in out-of-equilibrium scenarios. 
As quantum technologies continue to advance, our results could play a significant role in the development of efficient quantum thermal devices, providing a pathway to explore thermodynamic properties in regimes where traditional methods fall short.

\section{Acknowledgements}
This research was partially supported \emph{Centro Nazionale di Ricerca in High-Performance Computing, Big Data and Quantum Computing}, PNRR 4 2 1.4, CI CN00000013, CUP H23C22000360005.
A.P. would like to thank the Erasmus+ programme and the kind hospitality of the University of York.

\begin{thebibliography}{0}%
\makeatletter
\providecommand \@ifxundefined [1]{%
 \@ifx{#1\undefined}
}%
\providecommand \@ifnum [1]{%
 \ifnum #1\expandafter \@firstoftwo
 \else \expandafter \@secondoftwo
 \fi
}%
\providecommand \@ifx [1]{%
 \ifx #1\expandafter \@firstoftwo
 \else \expandafter \@secondoftwo
 \fi
}%
\providecommand \natexlab [1]{#1}%
\providecommand \enquote  [1]{``#1''}%
\providecommand \bibnamefont  [1]{#1}%
\providecommand \bibfnamefont [1]{#1}%
\providecommand \citenamefont [1]{#1}%
\providecommand \href@noop [0]{\@secondoftwo}%
\providecommand \href [0]{\begingroup \@sanitize@url \@href}%
\providecommand \@href[1]{\@@startlink{#1}\@@href}%
\providecommand \@@href[1]{\endgroup#1\@@endlink}%
\providecommand \@sanitize@url [0]{\catcode `\\12\catcode `\$12\catcode `\&12\catcode `\#12\catcode `\^12\catcode `\_12\catcode `\%12\relax}%
\providecommand \@@startlink[1]{}%
\providecommand \@@endlink[0]{}%
\providecommand \url  [0]{\begingroup\@sanitize@url \@url }%
\providecommand \@url [1]{\endgroup\@href {#1}{\urlprefix }}%
\providecommand \urlprefix  [0]{URL }%
\providecommand \Eprint [0]{\href }%
\providecommand \doibase [0]{https://doi.org/}%
\providecommand \selectlanguage [0]{\@gobble}%
\providecommand \bibinfo  [0]{\@secondoftwo}%
\providecommand \bibfield  [0]{\@secondoftwo}%
\providecommand \translation [1]{[#1]}%
\providecommand \BibitemOpen [0]{}%
\providecommand \bibitemStop [0]{}%
\providecommand \bibitemNoStop [0]{.\EOS\space}%
\providecommand \EOS [0]{\spacefactor3000\relax}%
\providecommand \BibitemShut  [1]{\csname bibitem#1\endcsname}%
\let\auto@bib@innerbib\@empty
\end{thebibliography}%


\begin{thebibliography}{99}
\bibitem{hohenberg1964density}
P.~Hohenberg and W.~Kohn, 
\textit{Inhomogeneous electron gas}, 
\href{https://doi.org/10.1103/PhysRev.136.B864}{Phys. Rev. \textbf{136}, B864 (1964)}.
\bibitem[Kohn and Sham(1965)]{kohn1965self}
W.~Kohn and L.~J.~Sham, 
\textit{Self-consistent equations including exchange and correlation effects}, 
\href{https://doi.org/10.1103/PhysRev.140.A1133}{Phys. Rev. \textbf{140}, A1133 (1965)}.

\bibitem{runge1984density}
E.~Runge and E.~K.~U.~Gross, 
\textit{Density-functional theory for time-dependent systems}, 
\href{https://doi.org/10.1103/PhysRevLett.52.997}{Phys. Rev. Lett. \textbf{52}, 997 (1984)}.

\bibitem{ullrich:2012}
C.~A.~Ullrich, 
\textit{Time-Dependent Density-Functional Theory}, 
\href{https://doi.org/10.1093/acprof:oso/9780199563029.001.0001}{Oxford University Press, 2012}.

\bibitem{mermin1965thermal}
N.~D.~Mermin, 
\textit{Thermal properties of the inhomogeneous electron gas}, 
\href{https://doi.org/10.1103/PhysRev.137.A1441}{Phys. Rev. \textbf{137}, A1441 (1965)}.

\bibitem{BSGP16}
K.~Burke, J.~C.~Smith, P.~E.~Grabowski, and A.~Pribram-Jones, 
\textit{Exact conditions on the temperature dependence of density functionals}, 
\href{https://doi.org/10.1103/PhysRevB.93.195132}{Phys. Rev. B \textbf{93}, 195132 (2016)}.

\bibitem{PGB16}
A.~Pribram-Jones, P.~E.~Grabowski, and K.~Burke, 
\textit{Thermal density functional theory: Time-dependent linear response and approximate functionals from the fluctuation-dissipation theorem}, 
\href{https://doi.org/10.1103/PhysRevLett.116.233001}{Phys. Rev. Lett. \textbf{116}, 233001 (2016)}.

\bibitem{SPB16}
J.~C.~Smith, A.~Pribram-Jones, and K.~Burke, 
\textit{Exact thermal density functional theory for a model system: Correlation components and accuracy of the zero-temperature exchange-correlation approximation}, 
\href{https://doi.org/10.1103/PhysRevB.93.245131}{Phys. Rev. B \textbf{93}, 245131 (2016)}.

\bibitem{YTPB14}
Z.-h.~Yang, J.~R.~Trail, A.~Pribram-Jones, K.~Burke, R.~J.~Needs, and C.~A.~Ullrich, 
\textit{Exact and approximate Kohn-Sham potentials in ensemble density-functional theory}, 
\href{https://doi.org/10.1103/PhysRevA.90.042501}{Phys. Rev. A \textbf{90}, 042501 (2014)}.

\bibitem{PPFS11}
S.~Pittalis, C.~R.~Proetto, A.~Floris, A.~Sanna, C.~Bersier, K.~Burke, and E.~K.~U.~Gross, 
\textit{Exact conditions in finite temperature density functional theory}, 
\href{https://doi.org/10.1103/PhysRevLett.107.163001}{Phys. Rev. Lett. \textbf{107}, 163001 (2011)}.

\bibitem{PhysRevB.105.235114}
P.~Hollebon and T.~Sjostrom, 
\textit{Hybrid Kohn-Sham+Thomas-Fermi scheme for high-temperature density functional theory}, 
\href{https://doi.org/10.1103/PhysRevB.105.235114}{Phys. Rev. B \textbf{105}, 235114 (2022)}.

\bibitem{PhysRevLett.125.055002}
A.~J.~White and L.~A.~Collins, 
\textit{Fast and universal Kohn-Sham density functional theory algorithm for warm dense matter to hot dense plasma}, 
\href{https://doi.org/10.1103/PhysRevLett.125.055002}{Phys. Rev. Lett. \textbf{125}, 055002 (2020)}.

\bibitem{PhysRevB.94.241103}
S.~Kurth and G.~Stefanucci, 
\textit{Nonequilibrium Anderson model made simple with density functional theory}, 
\href{https://doi.org/10.1103/PhysRevB.94.241103}{Phys. Rev. B \textbf{94}, 241103 (2016)}.

\bibitem{Karasiev:2016}
V.~V.~Karasiev, L.~Calderín, and S.~B.~Trickey, 
\textit{Importance of finite-temperature exchange correlation for warm dense matter calculations}, 
\href{https://doi.org/10.1103/PhysRevE.93.063207}{Phys. Rev. E \textbf{93}, 063207 (2016)}.

\bibitem{Trickey:2014}
F.~Graziani, M.~P.~Desjarlais, R.~Redmer, and S.~B.~Trickey, 
\textit{Frontiers and Challenges in Warm Dense Matter}, 
\href{https://doi.org/10.1007/978-3-319-04912-0}{Springer (2014)}.

\bibitem{PhysRevLett.78.2690}
C.~Jarzynski, 
\textit{Nonequilibrium equality for free energy differences}, 
\href{https://doi.org/10.1103/PhysRevLett.78.2690}{Phys. Rev. Lett. \textbf{78}, 2690 (1997)}.

\bibitem{RevModPhys.81.1665}
M.~Esposito, U.~Harbola, and S.~Mukamel, 
\textit{Nonequilibrium fluctuations, fluctuation theorems, and counting statistics in quantum systems}, 
\href{https://doi.org/10.1103/RevModPhys.81.1665}{Rev. Mod. Phys. \textbf{81}, 1665 (2009)}.

\bibitem{PhysRevLett.113.140601}
T.~B. Batalhão, A.~M. Souza, L. Mazzola, R. Auccaise, R.~S. Sarthour, I.~S. Oliveira, J. Goold, G. De~Chiara, M. Paternostro, and R.~M. Serra, \textit{Experimental reconstruction of work distribution and study of fluctuation relations in a closed quantum system}, \href{https://doi.org/10.1103/PhysRevLett.113.140601}{Phys. Rev. Lett. \textbf{113}, 140601 (2014)}.

\bibitem{PhysRevLett.115.190601}
T.~B. Batalhão, A.~M. Souza, R.~S. Sarthour, I.~S. Oliveira, M. Paternostro, E. Lutz, and R.~M. Serra, \textit{Irreversibility and the arrow of time in a quenched quantum system}, \href{https://doi.org/10.1103/PhysRevLett.115.190601}{Phys. Rev. Lett. \textbf{115}, 190601 (2015)}.

\bibitem{campaioli2024colloquium}
F. Campaioli, S. Gherardini, J.~Q. Quach, M. Polini, and G.~M. Andolina, \textit{Colloquium: Quantum batteries}, \href{https://doi.org/10.1103/RevModPhys.96.031001}{Rev. Mod. Phys. \textbf{96}, 031001 (2024)}.

\bibitem{campisi2011quantum}
M. Campisi, P. Hänggi, and P. Talkner, \textit{Quantum fluctuation relations and the arrow of time}, \href{https://doi.org/10.1103/RevModPhys.83.771}{Rev. Mod. Phys. \textbf{83}, 771 (2011)}.

\bibitem{goold2016role}
J. Goold, M. Huber, A. Riera, L. Del~Rio, and P. Skrzypczyk, \textit{The role of quantum information in thermodynamics—a topical review}, \href{https://doi.org/10.1088/1751-8113/49/14/143001}{J. Phys. A: Math. Theor. \textbf{49}, 143001 (2016)}.

\bibitem{klatzow2019experimental}
J. Klatzow, J.~N. Becker, P.~M. Ledingham, C. Weinzetl, K.~T. Kaczmarek, D.~J. Saunders, J. Nunn, I.~A. Walmsley, R. Uzdin, and E. Poem, \textit{Experimental demonstration of quantum effects in the operation of microscopic heat engines}, \href{https://doi.org/10.1103/PhysRevLett.122.110601}{Phys. Rev. Lett. \textbf{122}, 110601 (2019)}.

\bibitem{perarnau2015extractable}
M. Perarnau-Llobet, K.~V. Hovhannisyan, M. Huber, P. Skrzypczyk, N. Brunner, and A. Acín, \textit{Extractable work from correlations}, \href{https://doi.org/10.1103/PhysRevX.5.041011}{Phys. Rev. X \textbf{5}, 041011 (2015)}.

\bibitem{korzekwa2016extraction}
K. Korzekwa, M. Lostaglio, J. Oppenheim, and D. Jennings, \textit{The extraction of work from quantum coherence}, \href{https://doi.org/10.1088/1367-2630/18/2/023045}{New J. Phys. \textbf{18}, 023045 (2016)}.

\bibitem{shi2022entanglement}
H.-L. Shi, S. Ding, Q.-K. Wan, X.-H. Wang, and W.-L. Yang, \textit{Entanglement, coherence, and extractable work in quantum batteries}, \href{https://doi.org/10.1103/PhysRevLett.129.130602}{Phys. Rev. Lett. \textbf{129}, 130602 (2022)}.

\bibitem{gour2022role}
G. Gour, \textit{Role of quantum coherence in thermodynamics}, \href{https://doi.org/10.1103/PRXQuantum.3.040323}{PRX Quantum \textbf{3}, 040323 (2022)}.

\bibitem{rodrigues2024nonequilibrium}
F.~L. Rodrigues and E. Lutz, \textit{Nonequilibrium thermodynamics of quantum coherence beyond linear response}, \href{https://doi.org/10.1038/s42005-024-01015-6}{Commun. Phys. \textbf{7}, 61 (2024)}.

\bibitem{onishchenko2024probing}
O. Onishchenko, G. Guarnieri, P. Rosillo-Rodes, D. Pijn, J. Hilder, U. Poschinger, M. Perarnau-Llobet, J. Eisert, and F. Schmidt-Kaler, \textit{Probing coherent quantum thermodynamics using a trapped ion}, \href{https://doi.org/10.1038/s41467-024-02983-w}{Nat. Commun. \textbf{15}, 6974 (2024)}.

\bibitem{PhysRevE.99.042105}
G. Francica, J. Goold, and F. Plastina, \textit{Role of coherence in the nonequilibrium thermodynamics of quantum systems}, \href{https://doi.org/10.1103/PhysRevE.99.042105}{Phys. Rev. E \textbf{99}, 042105 (2019)}.

\bibitem{PhysRevE.105.014101}
G. Francica, \textit{Class of quasiprobability distributions of work with initial quantum coherence}, \href{https://doi.org/10.1103/PhysRevE.105.014101}{Phys. Rev. E \textbf{105}, 014101 (2022)}.

\bibitem{PhysRevResearch.2.033279}
A.~D. Varizi, A.~P. Vieira, C. Cormick, R.~C. Drumond, and G.~T. Landi, \textit{Quantum coherence and criticality in irreversible work}, \href{https://doi.org/10.1103/PhysRevResearch.2.033279}{Phys. Rev. Res. \textbf{2}, 033279 (2020)}.

\bibitem{varizi2021contributions}
A.~D. Varizi, M.~A. Cipolla, M. Perarnau-Llobet, R.~C. Drumond, and G.~T. Landi,
\textit{Contributions from populations and coherences in non-equilibrium entropy production},
\href{https://doi.org/10.1088/1367-2630/abfe20}{New J. Phys. \textbf{23}, 063027 (2021)}.

\bibitem{PhysRevResearch.5.043104}
M. Herrera, J.~H. Reina, I. D'Amico, and R.~M. Serra,
\textit{Correlation-boosted quantum engine: A proof-of-principle demonstration},
\href{https://doi.org/10.1103/PhysRevResearch.5.043104}{Phys. Rev. Res. \textbf{5}, 043104 (2023)}.

\bibitem{zawadzki2024work}
K. Zawadzki, G.~A. Canella, V.~V. Fran\c{c}a, and I. D'Amico,
\textit{Work statistics and entanglement across the fermionic superfluid-insulator transition},
\href{https://doi.org/10.1002/qute.202300237}{Adv. Quantum Technol. \textbf{7}, 2300237 (2024)}.

\bibitem{francica2020quantum}
G. Francica, F.~C. Binder, G. Guarnieri, M.~T. Mitchison, J. Goold, and F. Plastina,
\textit{Quantum coherence and ergotropy},
\href{https://doi.org/10.1103/PhysRevLett.125.180603}{Phys. Rev. Lett. \textbf{125}, 180603 (2020)}.

\bibitem{francica2017daemonic}
G. Francica, J. Goold, F. Plastina, and M. Paternostro,
\textit{Daemonic ergotropy: Enhanced work extraction from quantum correlations},
\href{https://doi.org/10.1038/s41534-017-0012-8}{NPJ Quantum Inf. \textbf{3}, 12 (2017)}.

\bibitem{PRXQuantum.4.020353}
K. Ptaszy\'{n}ski and M. Esposito,
\textit{Quantum and classical contributions to entropy production in fermionic and bosonic gaussian systems},
\href{https://doi.org/10.1103/PRXQuantum.4.020353}{PRX Quantum \textbf{4}, 020353 (2023)}.

\bibitem{geier2022non}
K.~T. Geier and P. Hauke,
\textit{From non-hermitian linear response to dynamical correlations and fluctuation-dissipation relations in quantum many-body systems},
\href{https://doi.org/10.1103/PRXQuantum.3.030308}{PRX Quantum \textbf{3}, 030308 (2022)}.

\bibitem{hahn2023quantum}
D. Hahn, M. Dupont, M. Schmitt, D.~J. Luitz, and M. Bukov,
\textit{Quantum many-body Jarzynski equality and dissipative noise on a digital quantum computer},
\href{https://doi.org/10.1103/PhysRevX.13.041023}{Phys. Rev. X \textbf{13}, 041023 (2023)}.

\bibitem{zhang2024energy}
J.-W. Zhang, B. Wang, W.-F. Yuan, J.-C. Li, J.-T. Bu, G.-Y. Ding, W.-Q. Ding, L. Chen, F. Zhou, and M. Feng,
\textit{Energy-conversion device using a quantum engine with the work medium of two-atom entanglement},
\href{https://doi.org/10.1103/PhysRevLett.132.180401}{Phys. Rev. Lett. \textbf{132}, 180401 (2024)}.

\bibitem{jussiau2023many}
{\'E}. Jussiau, L. Bresque, A. Auff\`{e}ves, K.~W. Murch, and A.~N. Jordan,
\textit{Many-body quantum vacuum fluctuation engines},
\href{https://doi.org/10.1103/PhysRevResearch.5.033122}{Phys. Rev. Research \textbf{5}, 033122 (2023)}.

\bibitem{jaramillo2016quantum}
J. Jaramillo, M. Beau, and A. del Campo,
\textit{Quantum supremacy of many-particle thermal machines},
\href{https://doi.org/10.1088/1367-2630/18/7/075019}{New J. Phys. \textbf{18}, 075019 (2016)}.

\bibitem{PhysRevResearch.2.032062}
T. Denzler and E. Lutz,
\textit{Efficiency fluctuations of a quantum heat engine},
\href{https://doi.org/10.1103/PhysRevResearch.2.032062}{Phys. Rev. Res. \textbf{2}, 032062 (2020)}.

\bibitem{zawadzki2022approximating}
K. Zawadzki, A.~H. Skelt and I. D’Amico,
\textit{Approximating quantum thermodynamic properties using DFT},
\href{https://doi.org/10.1088/1361-648X/ac6648}{J. Phys.: Condens. Matter \textbf{34}, 274002 (2022)}

\bibitem{skelt2019many}
A.~H. Skelt, K. Zawadzki and I. D’Amico, \textit{Many-body effects on the thermodynamics of closed quantum systems}, \href{https://doi.org/10.1088/1751-8121/ab4fb6}{J. Phys. A: Math. Theor. \textbf{52}, 485304 (2019)}.


\bibitem{herrera2018melting}
M. Herrera, K. Zawadzki, and I. D’Amico, \textit{Melting a Hubbard dimer: benchmarks of `ALDA' for quantum thermodynamics}, \href{https://doi.org/10.1140/epjb/e2018-90186-5}{Eur. Phys. J. B \textbf{91}, 1 (2018)}.

\bibitem{herrera2017dft}
M. Herrera, R.~M. Serra, and I. D’Amico, \textit{DFT-inspired methods for quantum thermodynamics}, \href{https://doi.org/10.1038/s41598-017-04478-y}{Sci. Rep. \textbf{7}, 4655 (2017)}.

\bibitem{PhysRevA.74.052335}
L.-A. Wu, M.~S. Sarandy, D.~A. Lidar, and L.~J. Sham, \textit{Linking entanglement and quantum phase transitions via density-functional theory}, \href{https://doi.org/10.1103/PhysRevA.74.052335}{Phys. Rev. A \textbf{74}, 052335 (2006)}.

\bibitem{PhysRevB.52.2504}
K. Sch\"{o}nhammer, O. Gunnarsson, and R.~M. Noack, \textit{Density-functional theory on a lattice: Comparison with exact numerical results for a model with strongly correlated electrons}, \href{https://doi.org/10.1103/PhysRevB.52.2504}{Phys. Rev. B \textbf{52}, 2504 (1995)}.

\bibitem{coe2015uniqueness}
J. Coe, I. D'Amico, and V. Fran\c{c}a, \textit{Uniqueness of density-to-potential mapping for fermionic lattice systems}, \href{https://doi.org/10.1209/0295-5075/110/63001}{Europhys. Lett. \textbf{110}, 63001 (2015)}.

\bibitem{penz2021density}
M. Penz and R. van Leeuwen,
\textit{Density-functional theory on graphs},
\href{https://doi.org/10.1063/5.0074249}{J. Chem. Phys. \textbf{155}, (2021)}.

\bibitem{xu2022extensibility}
L. Xu, J. Mao, X. Gao, and Z. Liu,
\textit{Extensibility of Hohenberg-Kohn theorem to general quantum systems},
\href{https://doi.org/10.1002/qute.202200041}{Adv. Quantum Technol. \textbf{5}, 2200041 (2022)}.

\bibitem{francca2018testing}
V.~V. Fran\c{c}a, J.~P. Coe, and I. D’Amico,
\textit{Testing density-functional approximations on a lattice and the applicability of the related Hohenberg-Kohn-like theorem},
\href{https://doi.org/10.1038/s41598-017-18642-w}{Sci. Rep. \textbf{8}, 664 (2018)}.

\bibitem{dreizler2012density}
R.~M. Dreizler and E.~K. Gross,
\textit{Density functional theory: an approach to the quantum many-body problem},
Springer Berlin, Heidelberg (2012).

\bibitem{alcaraz2008finite}
F.~C. Alcaraz and M.~S. Sarandy,
\textit{Finite-size corrections to entanglement in quantum critical systems},
\href{https://doi.org/10.1103/PhysRevA.78.032319}{Phys. Rev. A \textbf{78}, 032319 (2008)}.

\bibitem{pons2020hellmann}
M. Pons, B. Juliá-Díaz, A. Polls, A. Rios, and I. Vidana,
\textit{The Hellmann–Feynman theorem at finite temperature},
\href{https://doi.org/10.1119/10.0001184}{Am. J. Phys. \textbf{88}, 503 (2020)}.

\bibitem{landi2021irreversible}
G.~T. Landi and M. Paternostro,
\textit{Irreversible entropy production: From classical to quantum},
\href{https://doi.org/10.1103/RevModPhys.93.035008}{Rev. Mod. Phys. \textbf{93}, 035008 (2021)}.

\bibitem{talkner2007fluctuation}
P. Talkner, E. Lutz, and P. Hänggi,
\textit{Fluctuation theorems: Work is not an observable},
\href{https://doi.org/10.1103/PhysRevE.75.050102}{Phys. Rev. E \textbf{75}, 050102 (2007)}.

\bibitem{crooks1999entropy}
G.~E. Crooks,
\textit{Entropy production fluctuation theorem and the nonequilibrium work relation for free energy differences},
\href{https://doi.org/10.1103/PhysRevE.60.2721}{Phys. Rev. E \textbf{60}, 2721 (1999)}.

\bibitem{deffner2010generalized}
S. Deffner and E. Lutz,
\textit{Generalized Clausius inequality for nonequilibrium quantum processes},
\href{https://doi.org/10.1103/PhysRevLett.105.170402}{Phys. Rev. Lett. \textbf{105}, 170402 (2010)}.

\bibitem{kawai2007dissipation}
R. Kawai, J.~M.~R. Parrondo, and C.~V. den Broeck,
\textit{Dissipation: The phase-space perspective},
\href{https://doi.org/10.1103/PhysRevLett.98.080602}{Phys. Rev. Lett. \textbf{98}, 080602 (2007)}.

\bibitem{messiah2014quantum}
A. Messiah,
\textit{Quantum Mechanics},
Courier Corporation, Mineola, NY (2014).

\bibitem{wood1991systematic}
R.~H. Wood, W.~C. Muhlbauer, and P.~T. Thompson,
\textit{Systematic errors in free energy perturbation calculations due to a finite sample of configuration space: Sample-size hysteresis},
\href{https://doi.org/10.1021/j100171a062}{J. Phys. Chem. \textbf{95}, 6670 (1991)}.

\bibitem{hendrix2001fast}
D. Hendrix and C. Jarzynski,
\textit{A ``fast growth'' method of computing free energy differences},
\href{https://doi.org/10.1021/jp010198v}{J. Phys. Chem. \textbf{114}, 5974 (2001)}.

\bibitem{miller2019work}
H.~J.~D. Miller, M. Scandi, J. Anders, and M. Perarnau-Llobet,
\textit{Work fluctuations in slow processes: Quantum signatures and optimal control},
\href{https://doi.org/10.1103/PhysRevLett.123.230603}{Phys. Rev. Lett. \textbf{123}, 230603 (2019)}.

\bibitem{scandi2020quantum}
M. Scandi, H.~J.~D. Miller, J. Anders, and M. Perarnau-Llobet,
\textit{Quantum work statistics close to equilibrium},
\href{https://doi.org/10.1103/PhysRevResearch.2.023377}{Phys. Rev. Res. \textbf{2}, 023377 (2020)}.

\bibitem{capelle2013density}
K. Capelle and V.~L. Campo~Jr,
\textit{Density functionals and model Hamiltonians: Pillars of many-particle physics},
\href{https://doi.org/10.1016/j.physrep.2013.03.001}{Phys. Rep. \textbf{528}, 91 (2013)}.

\bibitem{franca2008entanglement}
V.~V. Fran\c{c}a and K. Capelle,
\textit{Entanglement in spatially inhomogeneous many-fermion systems},
\href{https://doi.org/10.1103/PhysRevLett.100.070403}{Phys. Rev. Lett. \textbf{100}, 070403 (2008)}.

\bibitem{gorling1994exact}
A. G\"{o}rling and M. Levy,
\textit{Exact Kohn-Sham scheme based on perturbation theory},
\href{https://doi.org/10.1103/PhysRevA.50.196}{Phys. Rev. A \textbf{50}, 196 (1994)}.

\bibitem{gorling1993correlation}
A. G\"{o}rling and M. Levy,
\textit{Correlation-energy functional and its high-density limit obtained from a coupling-constant perturbation expansion},
\href{https://doi.org/10.1103/PhysRevB.47.13105}{Phys. Rev. B \textbf{47}, 13105 (1993)}.

\bibitem{PhysRevLett.84.4255}
S. Pratt,
\textit{Canonical and microcanonical calculations for Fermi systems},
\href{https://doi.org/10.1103/PhysRevLett.84.4255}{Phys. Rev. Lett. \textbf{84}, 4255 (2000)}.

\bibitem{borrmann1993recursion}
P. Borrmann and G. Franke,
\textit{Recursion formulas for quantum statistical partition functions},
\href{https://doi.org/10.1063/1.464439}{J. Chem. Phys. \textbf{98}, 2484 (1993)}.

\bibitem{PhysRevE.83.067701}
J.-C. Pain, F. Gilleron, and Q. Porcherot,
\textit{Generating functions for canonical systems of fermions},
\href{https://doi.org/10.1103/PhysRevE.83.067701}{Phys. Rev. E \textbf{83}, 067701 (2011)}.

\bibitem{PhysRevResearch.2.043206}
H. Barghathi, J. Yu, and A. Del Maestro,
\textit{Theory of noninteracting fermions and bosons in the canonical ensemble},
\href{https://doi.org/10.1103/PhysRevResearch.2.043206}{Phys. Rev. Res. \textbf{2}, 043206 (2020)}.

\bibitem{pribram2014thermal}
A. Pribram-Jones, S. Pittalis, E.~K.~U. Gross, and K. Burke,
\textit{Thermal density functional theory in context},
in \textit{Frontiers and Challenges in Warm Dense Matter},
edited by M. Schlanges and H.~R. Schmidt
\href{https://doi.org/10.1007/978-3-642-54041-4_2}.{(Springer, 2014), pp. 25--60.}

\bibitem{gutzwiller1963effect}
M.~C. Gutzwiller,
\textit{Effect of correlation on the ferromagnetism of transition metals},
\href{https://doi.org/10.1103/PhysRevLett.10.159}{Phys. Rev. Lett. \textbf{10}, 159 (1963)}.

\bibitem{hubbard1963electron}
J. Hubbard,
\textit{Electron correlations in narrow energy bands},
\href{https://doi.org/10.1098/rspa.1963.0204}{Proc. R. Soc. Lond. A \textbf{276}, 238 (1963)}.

\bibitem{kanamori1963electron}
J. Kanamori,
\textit{Electron correlation and ferromagnetism of transition metals},
\href{https://doi.org/10.1143/PTP.30.275}{Prog. Theor. Phys. \textbf{30}, 275 (1963)}.

\bibitem{PhysRevLett.114.080402}
S. Murmann, A. Bergschneider, V.~M. Klinkhamer, G. Z\"urn, T. Lompe, and S. Jochim,
\textit{Two fermions in a double well: Exploring a fundamental building block of the Hubbard model},
\href{https://doi.org/10.1103/PhysRevLett.114.080402}{Phys. Rev. Lett. \textbf{114}, 080402 (2015)}.

\bibitem{editorial2013hubbard}
Editorial,
\textit{The Hubbard model at half a century},
\href{https://doi.org/10.1038/nphys2707}{Nat. Phys. \textbf{9}, 523 (2013)}.

\bibitem{carrascal2015hubbard}
D. Carrascal, J. Ferrer, J.~C. Smith, and K. Burke,
\textit{The Hubbard dimer: A density functional case study of a many-body problem},
\href{https://doi.org/10.1088/0953-8984/27/39/393001}{J. Phys.: Condens. Matter \textbf{27}, 393001 (2015)}.

\bibitem{PhysRevA.92.013614}
V.~L. Campo,
\textit{Density-functional-theory approach to the thermodynamics of the harmonically confined one-dimensional Hubbard model},
\href{https://doi.org/10.1103/PhysRevA.92.013614}{Phys. Rev. A \textbf{92}, 013614 (2015)}.

\bibitem{PhysRevB.86.235139}
G. Xianlong, A.-H. Chen, I.~V. Tokatly, and S. Kurth,
\textit{Lattice density functional theory at finite temperature with strongly density-dependent exchange-correlation potentials},
\href{https://doi.org/10.1103/PhysRevB.86.235139}{Phys. Rev. B \textbf{86}, 235139 (2012)}.

\bibitem{PhysRevLett.107.216401}
G. Stefanucci and S. Kurth,
\textit{Towards a description of the Kondo effect using time-dependent density-functional theory},
\href{https://doi.org/10.1103/PhysRevLett.107.216401}{Phys. Rev. Lett. \textbf{107}, 216401 (2011)}.

\bibitem{PhysRevLett.106.050401}
I. D'Amico, J.~P. Coe, V.~V. Fran\c{c}a, and K. Capelle,
\textit{Quantum mechanics in metric space: Wave functions and their densities},
\href{https://doi.org/10.1103/PhysRevLett.106.050401}{Phys. Rev. Lett. \textbf{106}, 050401 (2011)}.

\bibitem{levy1979universal}
M. Levy,
\textit{Universal variational functionals of electron densities, first-order density matrices, and natural spin-orbitals and solution of the v-representability problem},
\href{https://doi.org/10.1073/pnas.76.12.6062}{Proc. Natl. Acad. Sci. U.S.A. \textbf{76}, 6062 (1979)}.

\bibitem{PhysRevA.31.1950}
T.-c. Li and P.-q. Tong,
\textit{Hohenberg-Kohn theorem for time-dependent ensembles},
\href{https://doi.org/10.1103/PhysRevA.31.1950}{Phys. Rev. A \textbf{31}, 1950 (1985)}.

\bibitem{PhysRevB.83.035127}
I.~V. Tokatly,
\textit{Time-dependent current density functional theory on a lattice},
\href{https://doi.org/10.1103/PhysRevB.83.035127}{Phys. Rev. B \textbf{83}, 035127 (2011)}.

\bibitem{PhysRevLett.101.166401}
C. Verdozzi,
\textit{Time-dependent density-functional theory and strongly correlated systems: Insight from numerical studies},
\href{https://doi.org/10.1103/PhysRevLett.101.166401}{Phys. Rev. Lett. \textbf{101}, 166401 (2008)}.

\end{thebibliography}

\begin{appendix}
\section{Unique mapping for closed quantum systems \label{app1}}
We present here the MHK theorem for closed quantum systems described by Hamiltonians of the form~(\ref{Hl}). 
Let us consider the following Hamiltonians:
\begin{equation}
\label{HA1}
\hat{\mathcal{H}} = \hat{\mathcal{H}}_{0} +\sum^{\mathcal{L}}_{i=1} \lambda_i \hat{A}_i =\hat{\mathcal{H}}[\{\lambda_i\}]
\end{equation}
and
\begin{equation}
\label{HA2}
\hat{\mathcal{H}}^{\prime}= \hat{\mathcal{H}}_{0} + \sum^{\mathcal{L}}_{i=1} \lambda^{\prime}_i \hat{A}_i=\hat{\mathcal{H}}[\{\lambda_i^{\prime}\}],
\end{equation}
respectively associated to the sets of external parameters $\{\lambda_i\}$ and $\{\lambda_i^{\prime}\}$ of fixed length $\mathcal{L}$, which differ by more than a site-independent constant. 
The free energy corresponding to $\hat{\mathcal{H}}$, as given by Eq.~(\ref{F(rho)}), is minimized by the Gibbs state $\hat{\rho}^{\msc{th}}_{\beta}=\hat{\rho}^{\msc{th}}_{\beta}[\{\lambda_i\}]$.
We assume that this state is unique.
Similarly, the free energy corresponding to $\hat{\mathcal{H}}$ is minimized by $\hat{\rho}^{\msc{th}\prime}_{\beta}=\hat{\rho}^{\msc{th}}_{\beta\prime}[\{\lambda_i^{\prime}\}]$.
We also assume that this state is unique.
Given the nontrivial choice for the two sets of external parameters, the two above introduced thermal states must e different: $\hat{\rho}^{\msc{th}\prime}_{\beta}\neq\hat{\rho}^{\msc{th}}_{\beta}$.
Suppose that both $\{\lambda_i\}$ and $\{\lambda_i^{\prime}\}$ result in the same set of mean values $\{a^\beta_i\}$.
The two Hamiltonians are simply related by:
\begin{equation}
\hat{\mathcal{H}}^{\prime} = \hat{\mathcal{H}} + \sum^{\mathcal{L}}_{i=1} (\lambda^{\prime}_i - \lambda_i)\hat{A}_i.
\end{equation}
Consequently, the equilibrium free energy corresponding to the Hamiltonian~(\ref{HA2}) can be expressed as:
\begin{align}
\mathcal{F}^{\prime}[\hat{\rho}^{\msc{th}\prime}_{\beta}] &= \Tr\bigg\{
\hat{\rho}^{\msc{th}\prime}_{\beta}\bigg[
\hat{\mathcal{H}} 
+ \sum^{\mathcal{L}}_{i=1} ( \lambda_i^{\prime} - \lambda_i)\hat{A}_i 
+ \frac{\ln{\hat{\rho}^{\msc{th}\prime}_{\beta}}}{\beta}\bigg]\bigg\}\notag\\
&= \mathcal{F}[\hat{\rho}^{\msc{th}\prime}_{\beta}] 
+\sum^{\mathcal{L}}_{i=1} (\lambda_i^{\prime}- \lambda_i)a^{\beta}_{i}.
\end{align}
By the minimization principle of the equilibrium free energy, we must have $\mathcal{F}[\hat{\rho}^{\msc{th}\prime}_{\beta}]>\mathcal{F}[\hat{\rho}^{\msc{th}}_{\beta}]$, which leads to the inequality
\begin{equation}
\label{ineq1}
\mathcal{F}^{\prime}[\hat{\rho}^{\msc{th}\prime}_{\beta}] > \mathcal{F}[\hat{\rho}^{\msc{th}}_{\beta}]+ \sum^{\mathcal{L}}_{i=1} (\lambda^{\prime}_i - \lambda_i )a^{\beta}_{i}
\end{equation}
This inequality remains valid when primed and unprimed quantities are interchanged, yielding:
\begin{equation}
\label{ineq2}
\mathcal{F}[\hat{\rho}^{\msc{th}}_\beta] > \mathcal{F}^{\prime}[\hat{\rho}^{\msc{th}\prime}_{\beta}] + \sum^{\mathcal{L}}_{i=1} (\lambda_i - \lambda_i^{\prime})a^{\beta}_i 
\end{equation}
As in the original formulation of the first HK theorem~\cite{hohenberg1964density}, Eqs.~(\ref{ineq1}) and~(\ref{ineq2}) lead to the absurd conclusion that $\mathcal{F}^{\prime}[\hat{\rho}^{\msc{th}\prime}_{\beta}]+\mathcal{F}[\hat{\rho}^{\msc{th}}_\beta]> \mathcal{F}[\hat{\rho}^{\msc{th}}_{\beta}]+\mathcal{F}^{\prime}[\hat{\rho}^{\msc{th}\prime}_{\beta}]$.
Therefore, one and only one set of nontrivial external parameters, $\{\lambda_i\}$, can result in a given set of mean values, $\{a^\beta_i\}$, provided that the equilibrium Gibbs state is unique.
As a corollary, since $\{a^\beta_i\}$ uniquely determines $\{\lambda_i\}$, which in turn determines $\hat{\rho}^{\msc{th}}_\beta$, the thermal state is also a functional of the set $\{a^\beta_i\}$. Specifically,
\begin{equation}
 \{\lambda_i\} \iff \{a^\beta_i\} \iff \hat{\rho}^{\msc{th}}_\beta \equiv \hat{\rho}^{\msc{th}}_\beta[\{a^\beta_i\}].
\label{main}
\end{equation}
As a final note, we observe that this proof is based on the variational principle satisfied by the equilibrium free energy. 
Similarly, one could use an approach based on the constrained-search technique~\cite{levy1979universal}, thus removing the hypothesis on the uniqueness of the Gibbs thermal state, as it has been done in a previous study conducted at zero temperature~\cite{PhysRevB.52.2504}.

\section{Average work and irreversible entropy productions for general finite-time  protocols\label{app2}}
The Runge-Gross theorem for statistical mixtures~\cite{PhysRevA.31.1950}  ensures that the temporal evolution of an initial thermal state $\hat{\rho}^{\msc{th}}_\beta$
is a functional of the out-of-equilibrium electron density $n_\tau(\mathbf{r})$, i.e.:
\begin{align}
    \hat{\rho}(\tau)=&\hat{\mathcal{U}}(\tau,0)\hat{\rho}^{\msc{th}}_\beta[n_0^\beta(\mathbf{r})]{\mathcal{U}}^\dagger(\tau,0) \notag\\
    =&\hat{\rho}_\tau[n_\tau(\mathbf{r}),n_0^\beta(\mathbf{r})].
    \label{rhont}
\end{align}
Previous studies~\cite{PhysRevB.83.035127,PhysRevLett.101.166401} have shown that for pure states, an analogous form of the Runge-Gross theorem holds for systems driven by the Hamiltonians~(\ref{Hl}). 
Assuming that the same proof can be extended to statistical mixtures, equation~(\ref{rhont}) becomes
\begin{align}
    \hat{\rho}(\tau)=&\hat{\mathcal{U}}(\tau,0)\hat{\rho}^{\msc{th}}_\beta[\{a^{\beta{\,}0}_i\}]{\mathcal{U}}^\dagger(\tau,0)\notag \\ 
    =&  \hat{\rho}_\tau[\{a^{\beta{\,}\tau}_i\},\{a^{\beta{\,}0}_i\}].
    \label{rhontl}
\end{align}
By defining the out-of-equilibrium free energy as in Eq.~(\ref{F(rho)}), 
the average work and the irreversible entropy production 
respectively become:
\begin{equation}
\label{wwww}
\braket{w}=\mathcal{F}[\hat{\rho}(\tau)]-\mathcal{F}[\hat{\rho}^{\msc{th}{\,}0}_\beta],
\end{equation}
with $\hat{\rho}^{\msc{th}{\,}0}_\beta\equiv\hat{\rho}^{\msc{th}}_\beta[\{a^{\beta{\,}0}_i\}]$, and
\begin{equation}
\braket{\mathcal{S}_{\msc{irr}}}=\beta\{\mathcal{F}[\hat{\rho}(\tau)]-\mathcal{F}[\hat{\rho}^{\msc{th}{\,}\tau}_\beta]\},
\label{wsss}
\end{equation}
with $\hat{\rho}^{\msc{th}{\,}\tau}_\beta\equiv\hat{\rho}^{\msc{th}}_\beta[\{a^{\beta{\,}\tau}_i\}]$.

Then, using Eq~(\ref{rhontl}), these two expressions can be rewritten as:
\begin{equation}
\braket{w}=\sum_i\{\mathcal{\lambda}^{\tau}_{i}a^{\tau}_i-\mathcal{\lambda}^{0}_{i}a^{\beta{\,}0}_i\}
+\{\Omega[\{a^\tau_i\}]-\Omega[\{a^{\beta{\,}0}_i\}]\}
\label{wla}
\end{equation}
and
\begin{align}
\braket{\mathcal{S}_{\msc{irr}}}=&\beta\sum_i\{\mathcal{\lambda}^{\tau}_{i}a^{\tau}_i-\mathcal{\lambda}^{0}_{i}a^{\beta{\,}0}_i\}\notag\\
&-\beta\{\Omega[\{a^\tau_i\}]-\Omega[\{a^{\beta{\,}\tau}_i\}]\},
\label{Sla}
\end{align}
where $\Omega[\{a^\tau_i\}]$ represents the universal part of the out-of-equilibrium free energy. 
As previously stated, this part is generally unknown, which significantly increases the complexity of the problem. 
However, suitable approximations can generally be found~\cite{zawadzki2022approximating,skelt2019many,herrera2018melting,herrera2017dft}, highly dependent on the regimes under consideration, ranging from adiabatic to nearly sudden-quench evolutions.
Indeed, in the  $\tau \rightarrow 0$ limit, Eqs.~(\ref{wla}) and~(\ref{Sla}) reduce to the expressions derived in sec~\ref{sub:sqp} for sudden-quench protocols.

\section{Derivation of the fluctuation-dissipation relations for infinitesimal sudden quenches \label{app3}}
We provide here a derivation of Eqs.~(\ref{w^2}) and~(\ref{w^22}), yielding the fluctuation-dissipation relations~(\ref{FDRcl}) and~(\ref{FDR2}) in the sudden quench regime.
We then consider a sudden quench protocol starting from a thermal state $\hat{\rho}^{\msc{th}}_\beta$, or more generally for an initial Gibbs state commuting with $\hat{\mathcal{H}}[\{\lambda^0_i\}]$.
Given the  characteristic function of work~(\ref{chiws}), it is straightforward to show that the 
$n$-th moment of the distribution can be written as:
\begin{equation}
\braket{w^n}= \Tr\{(\hat{\mathcal{H}}[\{\lambda^f_i\}]-\hat{\mathcal{H}}[\{\lambda^0_i\}])^n\hat{\rho}^{\msc{th}}_\beta[\{\lambda^0_i\}]\}
\end{equation}
Specifically, the second moment of the PDW in a system defined by the Hamiltonian~(\ref{Hl}), where the external work parameters switch instantaneously from $\{\lambda_i\}$ to $\{\lambda_i+\delta\lambda_i\}$, is given by:
\begin{equation}
\braket{w^2}=\sum_i\sum_j \delta \lambda_i \delta\lambda_j \Tr\{ \hat{A}_i\hat{A}_j\hat{\rho}^{\msc{th}}_\beta[\{\lambda^0_i\}]\}.
\label{w2gen}
\end{equation} 
At this point, we want to make explicit the dependence on the thermal densities $\{a^\beta_i\}$. 

First, we consider the case where $[\hat{\mathcal{H}}[\{\lambda^f_i\}],\hat{\mathcal{H}}[\{\lambda^0_i\}]]=0$. 
The HF theorem, see Eq.~(\ref{af}), ensures that 
\begin{equation}
\frac{\partial^2 \mathcal{F} }{\partial \lambda_j\partial \lambda_i}=\frac{\partial a^\beta_i }{\partial \lambda_j}
\label{app31}
\end{equation}
and
\begin{equation}
\frac{1}{\mathcal{Z}}\frac{\partial \mathcal{Z}}{\partial \lambda_j}=-\beta a^\beta_j.
\end{equation}
On the other hand, we have 
\begin{equation}
\frac{\partial \hat{\rho}^{\msc{th}}_\beta}{\partial\lambda_j}=-\beta\hat{A}_j \hat{\rho}^{\msc{th}}_\beta- \hat{\rho}^{\msc{th}}_\beta\frac{1}{\mathcal{Z}}\frac{\partial \mathcal{Z}}{\partial \lambda_j}
\label{drhodl}
\end{equation}
and
\begin{align}
\frac{\partial^2 \mathcal{F}}{\partial \lambda_j\partial \lambda_i}=&\Tr\bigg\{ \hat{A}_i\frac{\partial \hat{\rho}^{\msc{th}}_\beta}{\partial\lambda_j}\bigg\}\notag\\
=&-\beta\Tr\{ \hat{A}_i\hat{A}_j \hat{\rho}^{\msc{th}}_\beta\}-\Tr\{ \hat{A}_i\hat{\rho}^{\msc{th}}_\beta\}\frac{1}{\mathcal{Z}}\frac{\partial \mathcal{Z}}{\partial \lambda_j}\notag\\
=& -\beta\Tr\{ \hat{A}_i\hat{A}_j \hat{\rho}^{\msc{th}}_\beta\}+\beta a^\beta_ia^\beta_j.
\label{app32}
\end{align}
Comparing Eq.~(\ref{app31}) with Eq.~(\ref{app32}), we obtain: 
\begin{equation}
 \Tr\{ \hat{A}_i\hat{A}_j\hat{\rho}^{\msc{th}}_\beta[\{\lambda^0_i\}]\}=\beta a^\beta_ia^\beta_j-\frac{1}{\beta}\frac{\partial a^\beta_i }{\partial \lambda_j}.
\label{app33}
\end{equation}
Then, substituting Eq.~(\ref{app33}) into Eq.~(\ref{w2gen}), we get $\braket{w^2}$ 
as given in Eq.~(\ref{w^2}).

We now analyze the scenario where the protocol is non-commutative, and the relation~(\ref{drhodl})
is no longer valid. Here, we rather have:
\begin{align}
\frac{\partial \hat{\rho}^{\msc{th}}_\beta}{\partial\lambda_j}=& -\beta\hat{A}_j \hat{\rho}^{\msc{th}}_\beta- \hat{\rho}^{\msc{th}}_\beta\frac{1}{\mathcal{Z}}\frac{\partial \mathcal{Z}}{\partial \lambda_j}\notag\\
&+\sum^{\infty}_{n=2} \frac{(-1)^n}{n!}\beta^n \underbrace{[\hat{\mathcal{H}},[\hat{\mathcal{H}}\dots [\hat{\mathcal{H}},\hat{A}_j ]]]}_{n{\:}\mathrm{times}}\hat{\rho}^{\msc{th}}_\beta.
\end{align}
Therefore, Eq.~(\ref{w^2}) is corrected by an infinite series of expectation values over the initial thermal state, which are still functionals of the initial thermal densities:
\begin{equation}
\theta^{(n)}[\{a^{\beta{\,}0}_i\}]=\Tr\{\hat{A}_i\underbrace{[\hat{\mathcal{H}},[\hat{\mathcal{H}}\dots [\hat{\mathcal{H}},\hat{A}_j ]]]}_{n{\,}\mathrm{times}}\hat{\rho}^{\msc{th}}_\beta\}.
\end{equation}
Consequently, the second moment of the PDW reads:
\begin{equation}
\braket{w^2}=\braket{w^2}_c+ \Theta_2[\{a^{\beta{\,}0}_i\}],
\end{equation}
with 
\begin{align}
\Theta_2[\{a^{\beta{\,}0}_i\}]=&\sum_{i,j} \delta \lambda_i\delta \lambda_j \sum^{\infty}_{n=2}\frac{(-1)^n}{n!}\beta^{n-1} \\
&\times \Tr\{\hat{A}_i\underbrace{[\hat{\mathcal{H}},[\hat{\mathcal{H}}\dots [\hat{\mathcal{H}},\hat{A}_j ]]]}_{n{\,}\mathrm{times}}\hat{\rho}^{\msc{th}}_\beta\}\notag\\
&=\sum_i\sum_j \delta \lambda_i\delta \lambda_j \sum^{\infty}_{n=2}\frac{(-1)^n}{n!}\beta^{n-1}\theta^{(n)}[\{a^{\beta{\,}0}_i\}].\notag
\end{align}

\section{Hubbard model in a lattice with one site\label{app4}}
Let us consider the single-site Hubbard model whose Hamiltonian is
\begin{equation}
\hat{\mathcal{H}}=\msc{v}_0(\hat{n}_{\uparrow}+ \hat{n}_{\downarrow})+ U\hat{n}_{\uparrow}\hat{n}_{\downarrow}.
\label{Hub1}
\end{equation}
Eq.~(\ref{Hub1}) describes a closed system, with an interacting thermal density 
\begin{align}
\label{nbeta}
 n_ \beta=& \Tr\{\hat{n}\hat{\rho}^{\msc{th}}_\beta\} =\Tr\{(\hat{n}_{\uparrow}+\hat{n}_{\downarrow})\hat{\rho}^{\msc{th}}_\beta\} \notag \\
=&\frac{2[e^{-\beta \msc{v}_0}+e^{-\beta(2\msc{v}_0+U)}]}{\mathcal{Z}},
\end{align}
depending on the canonical partition function:  
\begin{equation}
\mathcal{Z}= \Tr\{e^{-\beta\hat{\mathcal{H}}}\}=1+2e^{-\beta \msc{v}_0}+e^{-\beta(2\msc{v}_0+U)}.
\end{equation}
Eq.~(\ref{nbeta}), defining the thermal occupations of the model, depends only on $\msc{v}_0$.
It can be  inverted explicitly to obtain
\begin{equation}
\msc{v}_0[n_\beta]= -U -\frac{1}{\beta}\ln{\mathcal{G}^{\beta}_{U}[n_\beta]},
\end{equation}
where
\begin{equation}
\mathcal{G}^{\beta}_{U}[n_\beta]= \frac{\delta n_\beta+\sqrt{\delta n_\beta^2+e^{-\beta U} (1-\delta n_\beta)^2}}{1-\delta n_\beta} 
\end{equation}
and $\delta n_\beta= n_\beta-1$.
With the external potential expressed as a functional of the density, we can also write the partition function as an explicit functional of the density: 
\begin{equation}
\mathcal{Z}[n_\beta]=1+e^{\beta U}(2\mathcal{G}^{\beta}_{U}[n_\beta]+\mathcal{G}^{\beta}_{U}[n_\beta]^2)
\end{equation}
Consequently, the non-interacting KS Hamiltonian for the single-site Hubbard model reads: 
\begin{equation}
\hat{\mathcal{H}}^{\msc{ks}}=V^{\msc{ks}}(\hat{n}_{\uparrow}+ \hat{n}_{\downarrow})=(\msc{v}_0+V^{\msc{h-xc}})(\hat{n}_{\uparrow}+ \hat{n}_{\downarrow}).
\end{equation}
As it has been done for the interacting single site it is possible to invert  the relation
\begin{equation}
n^{\msc{ks}}_ \beta= \Tr\{\hat{n}\hat{\rho}^{\msc{th}}_{\beta \msc{ks}}\}= \frac{2[e^{-\beta V^{\msc{ks}}}+e^{-2\beta(V^{\msc{ks}})}]}{1+2e^{-\beta V^{\msc{ks}}}+2e^{-2\beta V^{\msc{ks}}}},
\end{equation}
for the KS thermal density, and obtain:
\begin{equation}
V^{\msc{ks}}[n^{\msc{ks}}_\beta]= -\frac{1}{\beta}\ln{\frac{n^{\msc{ks}}_\beta}{2-n^{\msc{ks}}_\beta}}
\end{equation}
By forcing $n^{\msc{ks}}_\beta = n_\beta$, the defining expression of the H-XC potential can be written as:
\begin{equation}
V^{\msc{h-xc}}[n_\beta]= U+\frac{1}{\beta}\ln{\Gamma^{\beta}_{U}[n_\beta]},
\end{equation}
where:
\begin{equation}
\Gamma^{\beta}_{U}[n_\beta]= \frac{\delta n_\beta+\sqrt{\delta n_\beta^2+e^{-\beta U} (1-\delta n_\beta^2)}}{n^{\msc{ks}}_\beta}
\label{GambU}
\end{equation}
\end{appendix}

\end{document}